\documentclass[mksc,unblindrev]{informs3}
\OneAndAHalfSpacedXI 

\usepackage{amssymb,amsmath,amsxtra,amstext,xfrac,mathtools}
\usepackage{graphicx,ccaption,booktabs,enumerate,paralist,mdwlist}
\usepackage{dsfont,verbatim,latexsym}
\usepackage{bbm}
\usepackage{bm}
\usepackage{longtable,multirow,threeparttable}
\usepackage{float}
\usepackage{eurosym}
\usepackage[english]{babel}
\usepackage[utf8]{inputenc}
\usepackage{fancyhdr}
\usepackage{color}
\usepackage{booktabs}
\usepackage{amssymb}
\usepackage{caption}
\usepackage{subcaption}

\DeclareMathOperator*{\esssup}{ess\,sup}
\usepackage{algorithm,algorithmic}
\usepackage{xcolor}
\usepackage{xurl}
\usepackage[colorlinks=TRUE,citecolor=MyDarkBlue,urlcolor=MyDarkBlue,linkcolor=MyDarkBlue]{hyperref}
\definecolor{MyDarkBlue}{RGB}{158,0,0}


\usepackage{natbib}
 \bibpunct[, ]{(}{)}{,}{a}{}{,}%
 \def\bibfont{\footnotesize}%

\TheoremsNumberedThrough    
\ECRepeatTheorems
\EquationsNumberedThrough 
\newcommand{\R}{\mathbb{R}}

\newcommand{\E}{\mathbb{E}}

\newcommand{\bmx}{\bm{x}}
\newcommand{\bmX}{\bm{X}}

\newcommand{\bmtheta}{\bm{\theta}}

\newcommand{\bbP}{\mathbb{P}}

\newcommand{\bbE}{\mathbb{E}}

\newcommand{\calZ}{\mathcal{Z}}
\newcommand{\calF}{\mathcal{F}}

\newcommand{\calD}{\mathcal{D}}
\newcommand{\calG}{\mathcal{G}}
\newcommand{\calN}{\mathcal{N}}
\newcommand{\calX}{\mathcal{X}}

\theoremstyle{plain}

\newcommand{\bc}[1]{\left\{#1\right\}}

\usepackage{color-edits}
\definecolor{redorange}{RGB}{255, 68, 51}
\addauthor[Ruohan]{rz}{redorange}

\renewcommand{\theARTICLETOP}{}

\begin{document}
\TITLE{Personalized Policy Learning through Discrete Experimentation: Theory and Empirical Evidence}
\RUNTITLE{Personalized Policy Learning through Discrete Experimentation}
\RUNAUTHOR{Zhang et al.}

\ARTICLEAUTHORS{
\AUTHOR{Zhiqi Zhang$^1$, Zhiyu Zeng$^2$, Ruohan Zhan$^3$, Dennis J. Zhang$^1$}
\AFF{
$^1$ Washington University in St. Louis, St. Louis, MO\\
$^2$ Shanghai Jiao Tong University, Shanghai, Shanghai, China\\
$^3$ University College London, London,UK\\
z.zhiqi@wustl.edu, zhiyu.zeng@sjtu.edu.cn, ruohan.zhan@ucl.ac.uk, denniszhang@wustl.edu
}
}

\ABSTRACT{ 
Randomized Controlled Trials (RCTs), or A/B testing, have become the gold standard for optimizing various operational policies on online platforms. However, RCTs on these platforms typically cover a limited number of discrete treatment levels, while the platforms increasingly face complex operational challenges involving optimizing continuous variables, such as pricing and incentive programs. The current industry practice involves discretizing these continuous decision variables into several treatment levels and selecting the optimal discrete treatment level. This approach, however, often leads to suboptimal decisions as it cannot accurately extrapolate performance for untested treatment levels and fails to account for heterogeneity in treatment effects across user characteristics. This study addresses these limitations by developing a theoretically solid and empirically verified framework to learn personalized continuous policies based on high-dimensional user characteristics, using observations from an RCT with only a discrete set of treatment levels. Specifically, we introduce a deep learning for policy targeting (\textsf{DLPT}) framework that includes both personalized policy value estimation and personalized policy learning. We prove that our policy value estimators are asymptotically unbiased and consistent, and the learned policy achieves a $\sqrt{n}$-regret bound. We empirically validate our methods in collaboration with a leading social media platform to optimize incentive levels for content creation. Results demonstrate that our \textsf{DLPT} framework significantly outperforms existing benchmarks, achieving substantial improvements in both evaluating the value of policies for each user group and identifying the optimal personalized policy.

}
\maketitle

\section{Introduction}\label{sec:intro}

In the rapid evolution of digital platforms, it has become increasingly common for firms to make decisions by running randomized controlled trials (RCTs), also known as experimentation or A/B testing. This method randomly assigns users to either treatment groups (which receive new candidate decision variables as interventions) or a control group (which maintains the status quo), compares their outcomes, and selects the optimal levels for each user group. Tech giants frequently run thousands of online experiments simultaneously to fine-tune algorithms, optimize user interface designs, and tailor marketing messages \citep{kohavi2017surprising}. Recent advances in AI and machine learning have further enabled platforms to 
personalize interventions \citep{rafieian2023ai}.

One major limitation of running RCTs on these platforms is limited sample sizes: due to the finite number of users, these platforms cannot test too many treatment levels in a single RCT, nor can they accurately estimate personalized treatment effects for very small subsets of users. This limitation is less critical when platforms focus on comparing a small number of discrete policies, such as a newly updated recommender system versus the original one \citep{feldman2022customer}. However, challenges arise when decision variables are continuous and policies need to be personalized, a scenario that is increasingly common in practice. For instance, e-commerce platforms like Alibaba frequently adjust their pricing strategies to optimize sales and profits \citep{zhang2020long}. 

When decision variables are continuous or have too many levels, current industry practice typically involves discretizing the treatment into several distinct levels, randomly assigning each level to a user group, and selecting the best-performing one. For example, ride-sharing platforms may experiment with different incentive levels for drivers, selecting the level that delivers the highest return on investment \citep{christensen2023demand}, while also testing various price levels during peak demand periods to assess customer price elasticity \citep{cohen2016using}. Similarly, streaming platforms may experiment with different discount durations—offering lower per-period costs for longer subscription commitments—and choose the option that maximizes user engagement \citep{tian2020optimizing}.

While such industry practices can help identify the best treatment levels among those tested, discretization limits the range of decision variables evaluated and prevents extrapolation to treatment levels outside the experimented set. As a result, the selected treatment from the experiment may be significantly outperformed by the optimal decision within the full continuous space. Moreover, relying solely on simple comparisons from A/B test results overlooks complex interactions between treatments and high-dimensional user characteristics, often resulting in suboptimal and non-personalized policies. This highlights the need for developing methods to identify optimal personalized policies using observations from only a discrete set of treatment levels in an RCT.

Therefore, in this paper, we aim to address the following research questions. First, how can we accurately estimate policy value over a continuous treatment space, conditioned on high-dimensional features, when observing only a discrete number of experimental arms on this continuous variable? Furthermore, how can we derive the optimal personalized policy for each feature combination? Our goal is to develop a theoretical framework with provable guarantees and rigorously validate our methods through real-world RCTs and applications.

Theoretically, we propose a Deep Learning framework for Policy Targeting, referred to as {\sf DLPT}, to estimate policy values conditional on high-dimensional personalized features and to identify optimal personalized policies using only experimental data with discrete treatment levels. Specifically, our model assumes a semi-parametric data generation process, where the outcome variable depends  parametrically on  the continuous treatment variables but non-parametrically on the features of each data point (i.e., the context). The model first translates each data point's features into a set of nuisance parameters, which subsequently determines the parameters in the specified functional form of the continuous treatment variables. As a result, the model accommodates a wide range of continuous functions for treatments, making it particularly powerful for extrapolating treatment effects beyond the observed discrete treatment levels.

With this model, we implement a three-stage method for continuous policy evaluation and optimization. In the first stage, we estimate the nuisance parameters in our model via nonparametric methods like neural networks. In the second stage, we estimate the policy value using universal orthogonal score functions, leveraging the nuisance estimates from the first stage. This ensures that our estimator is asymptotically unbiased and semi-parametrically efficient. In the third stage, we learn the optimal personalized policy by optimizing the estimated policy values and  provide theoretical guarantees for the optimality of the learned policy. Particularly, our policy achieves a regret bound  of $O\left(r\left(1+\sqrt{\log(1/r)}\right)\sqrt{d/n} + r\sqrt{\log(1/\delta)/n}\right)$,
where $r$ specifies the distance between any two policies in the policy class $\Pi$ in terms of the estimated policy value, and $d$ represents the complexity of the policy class $\Pi$. Notably, our results demonstrate that the learned policy achieves a minimax optimal regret rate of $\sqrt{n}$ within a fixed policy class.

We then validate our framework empirically in collaboration with one of the leading social media platforms worldwide, referred to as ``Platform O." Similar to Instagram, users on Platform O can both create and consume content, and the platform’s revenue is primarily driven by advertisements viewed by users. The platform is particularly interested in improving user-generated content production, which in turn enhances user engagement and increases future ad revenue. In partnership with this platform, we conducted a large-scale RCT involving 12.3\% of its users from August 23 to August 29, 2021. The experiment aimed to encourage users to create new videos by offering monetary rewards. Four different levels of monetary rewards were tested, resulting in four treatment groups and one control group (which didn't receive the monetary incentives). Using this experiment, we aim to evaluate our framework's performance on both policy value estimation and policy evaluation,  benchmarked with a set of frameworks from the literature and industry practices.

Our empirical evaluation comprises two components. First, we assess the performance of \textsf{DLPT} against a set of benchmark methods using experimental data.  We begin by measuring the accuracy of Average Treatment Effect (ATE) estimates: for each of four discrete monetary‐reward levels, we withhold that level from training, estimate its effect on users’ video‐upload behavior, and compare to the true ATE.  \textsf{DLPT} achieves a mean absolute percentage error (MAPE) of 2.25\%, whereas benchmark MAPEs span from 7.18\% to 43.59\%, demonstrating substantially greater precision in recovering out‐of‐sample ATEs.  Next, we evaluate policy‐learning regret within the finite observed policy class (the five experimental reward levels).  By concealing half of the user subgroups during training and computing regret on these held‐out groups, we find that \textsf{DLPT} attains the lowest mean percentage regret (MPR) of 2.22\%, compared with benchmark MPRs ranging from 6.32\% to 10.05\%.

Second, we evaluate the ability of these frameworks to identify the personalized optimal incentive policies beyond the finite observed policy class.  We construct semi–synthetic data by fitting a fully connected neural network (FNN) to the observed treatments and user covariates; this FNN then serves as the “ground truth” data‐generating process.  Policy performance is evaluated across four classes: (i) the Finite Observed class (restricted to the five experimental levels), (ii) a Discretized Continuous class, (iii) a Linear class, and (iv) a DNN class.  Across all four classes, \textsf{DLPT} consistently outperforms the benchmarks, achieving MPRs between 0.01\% and 19.69\%, whereas benchmark MPRs lie between 0.66\% and 34.75\%.  Within each policy class, the policy learned by \textsf{DLPT} yields the lowest regret, underscoring its flexibility and superior capacity to recover optimal incentive policies.

Our study makes several pivotal contributions to both academic research and practical applications in experimentation and platform analytics. From a methodological perspective,  we are the first, to our best knowledge, to propose a framework for estimating policy values and learning personalized policy over a continuous treatment variable when only observing discrete sets of treatment levels. The proposed framework and estimation procedure guarantee a $\sqrt{n}$ regret bound for the learned policy. In  real-world settings, we have validated our method in collaboration with a major online platform. The results confirm that our framework not only upholds strong theoretical promises but also excels in practical deployments, effectively handling extensive user features and nuanced treatment effects. More broadly, we develop our method for  post-experiment analysis, allowing for seamless integration into existing experimental platforms. This facilitates its adoption by practitioners who can readily apply our approach to enhance decision-making processes without modifying their existing experimentation infrastructure.

The remainder of the paper is organized as follows. We review the related literature in Section~\ref{sec:review}. Section~\ref{sec:theo_framework} introduces the {\sf DLPT} framework, outlining our proposed three-stage method along with theoretical guarantees. In Section~\ref{sec:empirical}, we describe the empirical setting of a large-scale field experiment. Section~\ref{sec:eval_empirical} presents the empirical evaluation of the {\sf DLPT} framework using the experimental data. We further assess the performance of {\sf DLPT} through semi-synthetic studies in Section~\ref{sec:semi-syn}. Finally, we conclude the paper in Section~\ref{sec:conclusion}. 
All codes for synthetic studies can be accessed in the Github Repository.\footnote{See    \url{https://anonymous.4open.science/r/Policy-Learning-through-Discrete-Experimentation-Theory-and-Empirical-Evidence-0B0B}}



\section{Literature Review}\label{sec:review}
Our research builds on three key streams of literature: personalized pricing, policy learning and causal Machine Learning (ML), and experimentation on platforms.

\textbf{Personalized Pricing}.
Our paper connects to personalized pricing, a critically important and well-established topic that has garnered significant attention across the fields of operations, economics, and marketing. This research can broadly be categorized into two main types: (1) pure modeling studies, which develop sophisticated mathematical models to analyze broad market dynamics, such as the effects of personalized pricing on social welfare, market efficiency, consumer surplus, and producer profits \citep[e.g.,][]{aguirre2010monopoly,chen2020competitive,elmachtoub2021value,ke2024algorithmic}; and (2) applied economics studies, which rely on empirical analyses using observational or experimental data and to which our research contributes.

Applied economics studies in this stream focus on the consumer side of the market and aim to develop and implement personalized pricing strategies by empirically estimating demand curves at various price points.
\cite{smith2023optimal} implement several model specifications such as the Bayesian choice model and multinomial logit model (MNL) and use an inversed propensity weighted (IPW) estimator for profits. \cite{hitsch2024heterogeneous} use conditional average treatment effect (CATE) estimators to find optimal policies. Various models, including linear regression models \citep[e.g.,][]{aryal2024price}, random coefficient models \citep[e.g.,][]{rossi1996value,aribarg2021price}, and nonparametric regression models \citep[e.g.,][]{hausman1995nonparametric,chen2021nonparametric,wang2024nonparametric}, are also wildly used in demand estimation in applied research. In the meanwhile, ML techniques have increasingly been integrated into demand estimation within the field of personalized pricing, primarily due to the availability of richer and more comprehensive data sets \citep[e.g.,][]{bajari2015machine,qi2022offline,dube2023personalized,yoganarasimhan2023design,wei2024neural}. 

Table \ref{tab:literature} summarizes representative literature in applied research of personalized pricing with data used in the paper (whether experimental data or observational data), personalization capability, causal inference capability, optimized policy type (discrete or continuous), and validation of the methods proposed (using cross-validation, simulation data, or another experiment data). We contribute to this literature by proposing a double machine learning-based framework for personalized policy targeting that is capable of causal inference for discrete experiments, policy personalization over continuous policy range, and methods validation through real-world data as well as synthetic studies. 

\begin{table}[]
\centering\scriptsize\caption{Summary of Related Literature in Personalized Pricing and Policy Learning}
\begin{tabular}{cccccc}
    \hline 
    \hline \\[-1.8ex] 
Paper  & Data Used & Personalization & Causality & Policy Type & Validation \\ \hline \\[-1.8ex] 
\multicolumn{6}{c}{Personalized Pricing}  \\\hline \\[-1.8ex] 
\cite{rossi1996value}  &    Observational Data  &  \checkmark & NA &  Discrete      &   NA   \\   
\cite{chen2021nonparametric}& NA  &  \checkmark & NA &  Continuous      &   NA   \\   
\cite{qi2022offline} &  NA  &\checkmark  &\checkmark  &Continuous & Simulation Data\\ 
\cite{smith2023optimal}  & Observational Data  &  \checkmark & NA &  Discrete      &   Cross-Validation   \\ 
\cite{yoganarasimhan2023design}  & Experimental Data  &  \checkmark & \checkmark &  Discrete      &   Cross-Validation   \\ 
\cite{dube2023personalized}  &    Experimental Data  &  \checkmark & \checkmark &  Discrete      &   Another Experiment Data \\ 
\cite{hitsch2024heterogeneous} &  Experimental Data  &  \checkmark & \checkmark &  Discrete      &   Another Experiment  Data  \\ 
\cite{aryal2024price} & Observational Data  &  \checkmark & NA &  Discrete      &   NA   \\ \hline \\[-1.8ex] 

\multicolumn{6}{c}{Policy Learning}  \\\hline \\[-1.8ex] 
\cite{zhao2012estimating,zhao2015new}&  Experimental Data &\checkmark & NA &  Discrete &  Cross-Validation\\
\cite{chen2016personalized}& NA &\checkmark & NA &  Continuous & Observational Data\\ 
\cite{kallus2018policy}& NA &\checkmark & NA &  Continuous & Observational Data \\ 
\cite{chernozhukov2019semi}& NA &\checkmark & NA &  Continuous & Simulation Data\\
\cite{athey2021policy}& Observational Data  &  \checkmark & NA &  Discrete    &   NA   \\
\cite{zhou2023offline}& Observational Data &  \checkmark & NA & Discrete & NA\\
\hline 
    \hline \\[-1.8ex]                  
\end{tabular}\label{tab:literature}
\end{table}

\textbf{Policy Learning and Causal ML}. Our research speaks to policy learning, a dynamic field at the intersection of economics, medical research, operations, and ML, which has evolved significantly over the past few years. Conventionally, policy learning can be distinguished based on the nature of the treatment variables and divides into two categories: discrete and continuous variables. Discrete variables deal with decisions that have a set number of distinct options. \cite{athey2021policy} use doubly robust score to find optimal treatment allocation as a binary policy. \cite{zhou2023offline} propose a method for multi-action policy learning with historical data. \cite{xiong2024staggered} propose optimal experimental design for staggered rollouts. \cite{zhan2024policy} extend the policy learning to   adaptively collected data settings.
In medical research, there is a substantial body of literature dedicated to identifying optimal individualized treatment rule (ITR) using observational data \citep[e.g.,][]{zhao2012estimating,zhao2015new,pan2021improved}. 
Continuous variables optimization involves treatments that can vary without limitation. \cite{chen2016personalized} introduce a non-convex loss function designed to refine the individualized dose rule (IDR) problem, and the method requires randomized trials where candidate dose levels assigned to study subjects are randomly chosen from a continuous distribution within a safe range. \cite{kallus2018policy} propose a continuous-treatment evaluator with a kernel function to incorporate localized data from comparable treatments.Both \cite{chen2016personalized} and  \cite{kallus2018policy} validate their methods through warfarin dataset \citep[]{international2009estimation} with multiple levels of warfarin dosage. Our work builds on \cite{chernozhukov2019semi} where the authors focus on observational study and propose a semi-parametric efficient framework for off-policy evaluation and optimization with continuous action spaces. We underscore our contribution in providing our framework {\sf DLPT}, an easily implementable approach that can extrapolate from a limited number of discretized experiments to learn optimal personalized continuous treatments. Simultaneously, our method adeptly manages causal inference, enhancing its applicability and effectiveness in practical settings.

Our work also connects to the theory and application of double machine learning (DML) \citep[]{chernozhukov2018double}. By integrating machine learning with Neyman orthogonality, DML excels in estimating parameters of interest, even when faced with challenges such as regularization and overfitting biases in nuisance parameter estimation. \cite{farrell2020deep,farrell2021deep} have developed innovative nonasymptotic high-probability bounds for estimating nuisance parameters using deep neural networks (DNN). \cite{knaus2022double} utilizes DML to assess the effectiveness of four labor programs in Switzerland. \cite{ye2023deep} propose a deep learning-based causal inference framework for estimating treatment effects for combinatorial experiments, showcasing the integration of advanced machine learning techniques with causal analysis methodologies. We extend the existing DML framework to accommodate continuous policy learning settings. This extension leverages the strengths of machine learning in managing high-dimensional individual data. Our enhanced framework has been validated through the use of real-world experimental data alongside extensive synthetic studies, demonstrating its effectiveness and robustness in practical applications.

\textbf{Experimentation on Platforms}. Large-scale experiment analysis of platforms is typically segmented into pre-experiment studies and post hoc experiment analysis. Pre-experiment studies primarily concentrate on the design of the experiment. \cite{johari2022experimental} analyze the bias of estimators on different sides (customer-side and listing-side) of randomization for experiments on online platforms. \cite{ye2023cold} propose a two-sided randomization experiment to evaluate the effectiveness of new algorithms. 
Additionally, the field is seeing advances in adaptive sequential experiments, where user information is incorporated as it becomes available \citep[e.g.,][]{gur2022adaptive,zhao2024pigeonhole}, leading to updated treatment assignments \citep[e.g.,][]{xiong2023optimal} and optimized experimental strategies such as multiarm-bandit experimental design \citep[e.g.,][]{simchi2023multi}. There is also growing interest in addressing experimental interference, which can skew results in large-scale platforms. \cite{wager2021experimenting} tackle this by assuming structured interference and using mean-field models to account for it effectively.  \cite{bojinov2023design} derives the optimal design of switchback experiments considering the carry-over effect. \cite{holtz2024reducing} use cluster randomization to deal with possible interference issues. \cite{zhan2024estimating} apply structured choice model to address seller-side competition in recommendation phase. Recent methodological advances have also introduced augmented difference-in-differences methods for causal inference \citep{li2023augmented}.

In contrast, post hoc experiment analysis offers a wide range of applications, enabling platforms to perform causal inference based on observed outcomes of discrete treatments, thereby supporting more informed and effective decision-making.  For instance,\cite{yu2022delay} document the impact of waiting time on users abandonment on a ride-sharing platform. \cite{zeng2023impact} investigate the causal effects of social nudges on content creation through a large-scale experiment. Building upon this body of work, our contribution extends beyond traditional causal inference by offering a novel extrapolation from existing experiments. Our framework {\sf DLPT} is tailored for discrete experiment settings and requires only mild assumptions for the identification of nuisance parameters. It enhances operational flexibility significantly by enabling not only causal inference for current treatment arms but also personalized continuous policy learning. This approach allows for valid extrapolation without the need for conducting complex, multi-arm A/B tests, simplifying the experimental process while broadening the scope of actionable insights.

\section{Theoretical Framework}\label{sec:theo_framework}


In this section, we introduce our theoretical framework,  {\sf DLPT}, for personalized policy learning. The framework first extrapolates the causal effects of discrete experimented treatments to the entire continuous space, and then find the optimal personalized continuous-treatment policy based on the extrapolation. We start with formalizing the problem and then propose  our method accompanied with its theoretical guarantees.  


\textbf{Problem Setup.\footnote{
\textit{On notations:} Throughout the paper, vectors and matrices are denoted in boldface. Vectors are treated as column vectors, and $\bm{v}'$ denotes the transpose of vector $\bm{v}$. Random variables are represented by capital letters, and their realizations by lowercase letters.}} 
A platform conducts a randomized experiment to determine the optimal treatment level within a continuous treatment space \( \mathcal{T} = [0, \bar{T}] \subset \mathbb{R} \), where \( \bar{T} < \infty \), with the goal of maximizing an outcome of interest (e.g., revenue). Due to operational constraints, the platform can only implement \( m \) discrete treatment levels, \( \{ t_1, t_2, \ldots, t_m \} \subset \mathcal{T} \), during the experiment. Let \( T \) denote the treatment level assigned to a user. The platform observes the assigned treatment level \( T \), as well as high-dimensional and bounded pretreatment covariates \( \bm{X} \in \mathcal{X} \in \mathbb{R}^{d_{\bm X}} \). Conditional on covariates, each treatment level is assigned with equal probability,\footnote{Our method does not require uniform treatment assignment; this assumption is made purely for clarity of exposition.} i.e., $
\mathbb{P}(T = t_j \mid \bm{X} = \bm{x}) = \frac{1}{m}, \quad \forall j \in \{1, \ldots, m\}.$
After treatment assignment, the user reveals an outcome \( Y \in \mathcal{Y} \subset \mathbb{R} \), which the platform can observe. Without loss of generality, we focus on binary outcomes \( Y \in \{0,1\} \); our results extend directly to general bounded outcomes.

Following \cite{farrell2020deep}, we assume that the underlying data generation process (DGP) has a semi-parametric specification such that 
\begin{equation}\label{eq:dgp assumption}
    \mathbb{E}[Y | \bm X = \bm x, T= t]=G(\bm{\theta}^*(\bm{x}),t).
\end{equation}
Above, the link function $G(\cdot)$ is parametrically specified a priori, which can vary from linear to non-linear forms, such as sigmoid, depending on the specific domain knowledge and hypotheses stipulated by the researchers.
The nonparametric part, $\bm{\theta}^*(\cdot)$, is typically known as ``nuisance parameter'' in causal inference literature \citep{newey1994asymptotic}, which  is designed to be highly adaptable---without parametric restrictions---to encapsulate a summary of user characteristics and their interactions with the treatment level $t$. This  $\bm{\theta}^*(\cdot)$ shall be learned using deep neural networks (DNNs) in our empirical settings, which afford the flexibility needed to capture complex, non-linear relationships that may exist between user attributes and treatment effects. 

In our analysis, we assume a specific  sigmoid-polynomial specification for $G(\cdot)$ as the DGP. Accordingly, specification \eqref{eq:dgp assumption} is detailed as follows:
\begin{equation}\label{eq:DGP_ass}
    \mathbb{E}[Y | \bm X = \bm x, T= t]=G(\bm{\theta}_K^*(\bm{x}),t) = \frac{1}{1+\exp(-(\bm\theta_{K}^*(\bm x)'\tilde{\bm T}_K))},
\end{equation}
where \( \tilde{\bm{T}}_K(t) = (1, t, t^2, \ldots, t^K)' \in \mathbb{R}^{K+1} \) denotes the degree-\( K \) polynomial feature vector constructed from the treatment variable \( t \). The coefficient vector \( \bm{\theta}_K^*(\bm{x}) \in \mathbb{R}^{K+1} \) varies with covariates \( \bm{x} \), and governs the influence of each polynomial term in the transformed treatment input. As to be discussed in Remark 1, the optimal $K$ is chosen as the number of discrete treatments minus $1$. \emph{Note that this functional form assumption for $G(\cdot)$ may be misspecified. However, even if the true data generation process is arbitrary, our specification above can offer strong approximation power. We provide further theoretical analysis in Section~\ref{sec:approximation_power}.}

\textbf{Policy.} A policy $\pi$ is a treatment assignment rule that maps from the covariate space $\mathcal{X}$ to the treatment space $\mathcal{T}$. For a given user with covariates \( \bm{x} \), the value \( \pi(\bm{x}) \) denotes the treatment level that the policy prescribes for that user based on her characteristics. Let $H(\cdot)$ denote the value function chosen by researchers or practitioners. For example, $H(\bm{\theta}_K^*(\bm{x}),\pi(\bm x)) = wG(\bm{\theta}_K^*(\bm{x}),\pi(\bm x))-c\pi(\bm x)$ represents net benefit as the profit gained from users outcome with weight $w$ minus policy cost with unit cost $c$. For a given policy $\pi$, we use $V_K(\pi)$ to denote its policy value, which represents the expected value when the treatment assignment follows policy $\pi$: 
\begin{equation}\label{eq:policy value}
    V_K(\pi):= \bbE[H(\bm{\theta}_K^*(\bm{x}),\pi(\bm x))],
\end{equation}
where the expectation is taken over the covariate distribution.

\textbf{Goal.}
 Given a pre-specified policy class $\Pi$ (such as linear policies or decision-tree policies) and the current semi-parametric DGP assumption $G(\bm \theta^*_K(\bm x),t)$, we aim to learn a policy $\hat{\pi}\in\Pi$ from the data, such that its policy value is maximized, or equivalently,  we aim to minimize its regret $R_K(\hat{\pi})$, which measures the policy's suboptimality as defined below:
\begin{equation}
    R_K(\hat{\pi}) =  V_K(\pi_K^*) - V_K(\hat{\pi}),
\end{equation}
where $\pi_K^*=\arg\max_{\pi\in\Pi}V_K(\pi)$ denotes the optimal \emph{in-class} policy.

\textbf{Three-Stage Policy Learning Method: {\sf DLPT}.} We propose a three-stage method to learn optimal policies, which we term as Deep Learning Policy Targeting ({\sf DLPT}). The key is to causally extrapolate the policy value of continuous treatment $T$ from discrete observed experiments.  Our method unfolds in the following stages: (i) estimating the nuisance parameter  $\bm{\theta}_K^*(\bm{x})$  via structured ML models; (ii) estimating counterfactual policy values  using Neyman orthogonal scores; and (iii) learning the optimal personalized policy within the specified policy class. We detail each stage in sections \ref{sec:nuisance parameter est}, \ref{sec:policy_value_est}, and \ref{sec:policy_learning} respectively and complement these procedures with theoretical guarantees for policy value estimation and policy learning. 

In later sections, we shall instantiate our  {\sf DLPT} method in the field. Specifically,  our empirical setting considers outcome $Y$ as a binary variable indicating user video uploads, the treatment $T$ as a continuous variable denoting monetary incentive offered to encourage user creation, and $\bm X$ as users' features (See Section \ref{sec:empirical_setting} for detailed empirical setting). Our goal is to learn a cost-effective policy for offering monetary incentives. 
Although this illustration is specific to our empirical application, our {\sf DLPT} method is designed to be flexible and can be customized to accommodate various continuous outcome variables. Let \( n \) denote the number of users involved in the experiment, indexed by \( i \in \{1, \ldots, n\} \). For each user \( i \), let \( \bm{X}_i \), \( T_i \), and \( Y_i \) denote her covariates, treatment level, and observed outcome, respectively. At the conclusion of the experiment, the platform collects \( n \) i.i.d. observations: $\{ \bm{Z}_i = (\bm{X}_i', T_i, Y_i)' \}_{i=1}^n.$

\subsection{Stage 1: Nuisance Parameter Estimation}\label{sec:nuisance parameter est}
In the first stage, we approximate the link function \( G(\cdot) \) using a structured DNN. Specifically, we employ a flexible feed-forward DNN, denoted by \( \hat{\bm{\theta}}(\bm{x}) \), to approximate the nuisance parameter \( \bm{\theta}_K^*(\bm{x}) \). Following our sigmoid-polynomial specification introduced in Equation~\eqref{eq:DGP_ass}, the interaction between the treatment variable \( t \) and the estimated covariate-dependent coefficients \( \hat{\bm{\theta}}(\bm{x}) \) is modeled in a structured and explicit form:
\begin{equation}\label{eq:DGP}
    \mathbb{E}[Y \mid \bm{X} = \bm{x}, T = t] = G(\hat{\bm{\theta}}(\bm{x}), t) = \frac{1}{1 + \exp\left( -\left( \hat{\theta}_0(\bm{x}) + \hat{\theta}_1(\bm{x}) t + \cdots + \hat{\theta}_K(\bm{x}) t^K \right) \right)},
\end{equation}
where \( \hat{\theta}_k(\bm{x}) \) denotes the coefficient corresponding to the \( k \)-th polynomial term of the treatment variable, for \( k = 0, \ldots, K \), as a function of user covariates \( \bm{x} \). The nuisance parameter estimator \( \hat{\bm{\theta}}(\cdot) \) is learned by solving the following empirical risk minimization (ERM) problem:
\begin{equation}
    \hat{\bm{\theta}}(\cdot) = \arg\min_{\tilde{\bm{\theta}} \in \bm{\Theta}}~\frac{1}{n} \sum_{i=1}^n \ell(Y_i, T_i, \tilde{\bm{\theta}}(\bm{X}_i)),
\end{equation}
where \( \ell(\cdot) \) is an appropriate loss function (e.g., squared loss for continuous outcomes or cross-entropy loss for binary outcomes), and \( \bm{\Theta} \) denotes the functional space of all possible feed-forward DNN under a chosen set of hyperparameters.

To achieve a sufficiently rapid convergence rate of the nuisance estimates---which is crucial for Stage 2 (policy value estimation) and Stage 3 (policy learning)---we establish certain regularity conditions.  
These conditions are commonly assumed in the literature with feed-forward neural neowrks \citep{farrell2020deep},  pertain to the continuity and cuvation of loss function $l(\cdot)$ (Assumption \ref{ass:loss function} in Appendix \ref{app:assumptions}), the joint distribution of $(\bm X',T,Y)'$ (Assumption \ref{ass:data bounded and nuisance smoonth}(a) in Appendix \ref{app:assumptions}), and DGP nuisance parameter $\bm \theta^*(\bm x)$ (Assumption \ref{ass:data bounded and nuisance smoonth}(b) in Appendix \ref{app:assumptions}). Given the structure of \( \tilde{\bm{T}}_K(t) = (1, t, t^2, \ldots, t^K)' \), we analogously define the policy-induced polynomial vector as \( \tilde{\bm{\pi}}_K(\bm{x}) = (1, \pi(\bm{x}), \pi(\bm{x})^2, \ldots, \pi(\bm{x})^K)' \).
Proposition \ref{prop: nuisance convergence_rate} presents the convergence rate for estimating nuisance parameters during Stage 1, with its proof deferred to   Appendix \ref{appendix:prop convergence rate proof}. This result is foundational to ensuring the reliability and efficiency of subsequent stages of {\sf DLPT} implementation.

\begin{proposition}[\sc Indentifiability and Convergence of Nuisance Estimate] \label{prop: nuisance convergence_rate}
 Suppose that  $\mathbb{E}[\tilde{\bm T}_K \tilde{\bm T}_K'|\bm X]$ is positive definite across all $\bm X$, and  that Assumption \ref{ass:loss function} and Assumption \ref{ass:data bounded and nuisance smoonth} in Appendix \ref{app:assumptions} hold,
\begin{itemize}
    \item [(a)]  The nuisance parameter function $\bm{\theta}_K^*(\bm{x})$ can be nonparametrically identified in DGP (\ref{eq:DGP}).
    \item [(b) \citep{farrell2020deep,farrell2021deep}] If the structured DNN has width $H=O(n^{{d_C}/{2(p+d_C)}}\log^2 n)$ and depth $L=O(\log n)$,  there exists a positive constant $C$ that depends on the fixed quantities in Assumption \ref{ass:data bounded and nuisance smoonth}, such that with probability at least $1-\exp(n^{-{d_C}/{(p+d_{C})}}\log^8 n)$, for $n$ large enough, the convergence of $\hat{\bm \theta}$ satisfies 
\begin{equation}
    \bbE[\|\hat{\bm\theta}(\bmx)-\bm\theta^*_K(\bmx)\|_{L_2(\bm X)}^2] = O\Big(n^{-\frac{p}{p+d_C}}\log^8(n)+\frac{\log\log n}{n}\Big),
\end{equation}
where $p$ characterizes the smoothness of $\bm\theta^*_K(\cdot)$, and $d_C$ is the dimension of continuous features in context.
\end{itemize}
\end{proposition}

\begin{remark}[CHOOSING OPTIMAL $K$]\label{remark:selectK}
The uniform positive definiteness condition for $\mathbb{E}[\tilde{\bm T}_K \tilde{\bm T}_K'|\bm X]$ is satisfied when each user is randomly assigned $m\geq K+1$ discrete treatment values, regardless of their covariates $\bm X$. As to be discussed in Section~\ref{sec:approximation_power}, the higher $K$ in Equation~\eqref{eq:DGP}, the better the approximation power our method attains. Therefore, if an experiment with $m$ treatment conditions is run, then the optimal $K$ should be $m-1$ when applying our method.
\end{remark}

\subsection{Stage 2: Policy Value Estimation}\label{sec:policy_value_est}
In the second stage of our framework, we estimate the policy value of each counterfactual policy~$\pi$. Recall that the policy value is defined as:
\begin{equation}
\label{eq:policy_value}
    V_K(\pi) := \mathbb{E}[H(\bm{\theta}_K^*(\bm{x}),\pi(\bm{x}))] = \mathbb{E}[wG(\bm{\theta}_K^*(\bm{x}),\pi(\bm{x}))-c\pi(\bm{x})],
\end{equation}
where $w$ represents the unit profit gained from users' engagement outcomes, and $c$ denotes the unit cost of the policy. This formulation captures the expected net benefit of implementing policy~$\pi$.

With the nuisance estimate $\hat{\bm \theta}(\cdot)$ obtained from Stage 1, a natural policy value estimator  is to plug $\hat{\bm \theta}(\cdot)$ into Equation \eqref{eq:policy_value}.
However, the estimation error of $\hat{\bm \theta}(\cdot)$ will propagate to the policy value estimation, and 
the error diminishing rate guaranteed in Proposition  \ref{prop: nuisance convergence_rate} will be not fast enough to guarantee valid inference and prepare for efficient policy learning.


In light of this, we follow \cite{farrell2020deep} and propose a semi-parametric policy value estimator by using orthogonalized score functions of $\hat{\bm \theta}(\cdot)$. 
Score functions with this orthogonality condition, or more precisely, universal (Neyman) orthogonality condition,  are typically referred to as influence functions in econometrics literature \citep{newey1994asymptotic,ichimura2022influence}. These influence functions exhibit first-order robustness to small variations in nuisance parameter estimation, by having error decay rate faster than $\sqrt{n}$. This robustness allows for semi-parametrically inference and  efficient policy learning, even if the convergence speed of the nuisance estimates is not sufficient to achieve semi-parametric efficiency itself. 
\begin{proposition}[\sc Influence Function]\label{prop:influence function}
Assume Assumptions \ref{ass:loss function}, \ref{ass:data bounded and nuisance smoonth}, and \ref{ass:influence function} in Appendix \ref{app:assumptions} hold. $\psi(\bm z,\pi;\hat{\bm\theta},{\bm\Lambda})-V_K(\pi)$ is an  influence function of $V_K(\pi)$, where $\psi(\bm z,\pi;\hat{\bm\theta},{\bm\Lambda})$ is referred to as score function and is defined as: 
\begin{equation}\label{eq:influence function}
    \psi(\bm z,\pi;\hat{\bm\theta},\bm\Lambda) = H(\hat{\bm\theta}(\bm x),\pi(\bm x)) - H_{\bm\theta}(\hat{\bm\theta}(\bm x), \pi(\bm x))\bm\Lambda(\bm x)^{-1} \ell_{\bm\theta}(y,  t, \hat{\bm\theta}(\bm x)),
\end{equation}
where $\bm z = (\bm x',t,y)’$ is the observed data sample, $\ell_{\bm\theta}$ is the gradient of $\ell$ with respect to $\bm\theta$, $\bm \Lambda(\bm x)=\bbE[\ell_{\bm\theta\bm\theta}(y,t,\bm\theta(\bm x))|\bm X = \bm x]$ as the expectation of second order derivative of $\ell$ with respect to $\bm\theta$, and $H_{\bm\theta}(\hat{\bm\theta}(\bm x),\pi(\bm x))$ is the gradient of $H$ with respect to $\theta$.
\end{proposition}
\begin{remark}[Explicit computation of nuisance parameter $\bm\Lambda(\bm x)$]
Unlike general scenarios explored in \cite{farrell2020deep}, our empirical setting allows for a direct computation of the nuisance parameter $\bm\Lambda(\bm x)$, obviating the need for its estimation. This simplification is possible because the distribution of the treatment level $t$, defined by the random treatment assignment mechanism, is known a priori. Depending on the loss function employed in Stage 1 of our analysis, the form of $\bm\Lambda(\bm x)$ varies as follows:


\[
\bm\Lambda(\bm x) = 
\begin{cases}
\mathbb{E}\left[G_{\bm \theta}(\hat{\bm\theta}(\bm x), t) G_{\bm \theta}(\hat{\bm\theta}(\bm x), t)' \mid \bm X = \bm x\right], & \text{if using MSE loss} \\
\mathbb{E}\left[G_{\bm \theta}(\hat{\bm\theta}(\bm x), t)\tilde{\bm T}_K' \mid \bm X = \bm x\right], & \text{if using Binary Cross Entropy loss}
\end{cases}
\]
\[
\text{where } G_{\bm \theta}(\hat{\bm\theta}(\bm x), t) = G(\hat{\bm\theta}(\bm x), t)(1 - G(\hat{\bm\theta}(\bm x), t))\tilde{\bm T}_K.
\]

\end{remark}

With the score function in \eqref{eq:influence function}, the below result shows that its derivative with respect to our nuisance parameters evaluated at the true parameter values is 0. In other words, it satisfies Neyman orthogonality.

\begin{proposition}[\sc Universal Orthogonality]\label{prop:orthogonality}
    The score function  $\psi(\bm z,\pi;\hat{\bm\theta},\bm\Lambda)$ defined in Proposition \ref{prop:influence function} for estimating policy value $V_K(\pi)$ is universal  orthogonal with respect to the nuisance parameter $\bm \theta$. That is,
    \begin{equation}
        \bbE[\nabla_{\bm \theta}\psi(\bm z,\pi;\bm \theta^*, \bm \Lambda)|\bm X = \bm x]=0.
    \end{equation}
\end{proposition}

Now we are ready to introduce our  policy value estimator $\hat{V_K}^{\sf DLPT}(\pi)$, based on the score function in Proposition \ref{prop:influence function}, composed with the sample-splitting estimation procedure. Specifically, we randomly partition the dataset \( S \), consisting of \( n \) i.i.d.~samples, evenly into two subsets \( S_1 \) and \( S_2 \). We first estimate the nuisance parameters \(\hat{\bm{\theta}}\) using subset \( S_1 \) following the procedure outlined in Stage 1. Subsequently, we utilize subset \( S_2 \) for policy evaluation in Stage 2.
\begin{gather}\label{eq:dr_estimator}
   \hat{V}_K^{\sf DLPT}(\pi)=\frac{1}{|S_2|}\sum_{i\in S_2} \psi(\bm Z_i,\pi;\hat{\bm \theta},\bm\Lambda)\\
    \hat{\Psi}_K^{\sf DLPT}(\pi)=\frac{1}{|S_2|}\sum_{i\in S_2}\bigg(\psi(\bm Z_i,\pi;\hat{\bm \theta},\bm\Lambda)-\hat{V}_K^{\sf DLPT}(\pi)\bigg)^2
\end{gather}

Proposition \ref{prop:asy normal}, adapted from Theorem 3 in  \cite{farrell2020deep},  presents the $\sqrt{n}$-asymptotic normality of our policy value estimator $\hat{V}_K^{\sf DLPT}(\pi)$ with the estimated variance $\hat{\Psi}_K^{\sf DLPT}(\pi)$, allowing for policy value inference and hypothesis testing.  It also shows that, with the Neyman orthogonality score function, our estimator is consistent and asymptotically normal.


\begin{proposition}[\sc Asymptotic Normality]\label{prop:asy normal}
    Assume Assumptions \ref{ass:loss function}, \ref{ass:data bounded and nuisance smoonth}, and \ref{ass:influence function} in Appendix \ref{app:assumptions} hold, and $\bm\Lambda(\bm x)$ is uniformly invertible. Also, the nuisance parameter estimator $\hat{\bm \theta}_1$ in Stage 1 obeys 
    $||\hat{\bm \theta}_1-\bm\theta^*_1||_{L_2(\bm X)} = o(n^{-1/4})$
    , which is verified by Proposition \ref{prop: nuisance convergence_rate}(b). Then
    \begin{equation}
        \sqrt{n} \hat{\Psi}_K^{\sf DLPT}(\pi)^{-1/2} (\hat{V}_K^{\sf DLPT}(\pi)- V_K(\pi))\rightarrow_d\mathcal{N}(0,1).
    \end{equation}
\end{proposition}

Proposition~\ref{prop:asy normal} shows that we can construct an  semi-parametric estimator for evaluating any personalized policy $\pi$ from the data and conduct valid inference.
Note that this estimator can be used to evaluate any policies, not limited to the observed levels in the experiment. With this estimator, we next discuss how to learn the optimal personalized policy from the data.

\subsection{Stage 3: Policy Learning}\label{sec:policy_learning}
 We denote the optimal policy derived from our framework as \(\hat{\pi}^{\Pi}_{DLPT}\), defined as the policy within a specified class \(\Pi\) that maximizes the expected policy value. Formally, this policy \(\hat{\pi}^{\Pi}_{DLPT}\) is obtained by solving the constrained empirical risk minimization (ERM) problem:
\begin{equation}
\label{eq:policy_learning}
    \hat{\pi}^{\Pi}_{DLPT} = \inf_{\pi\in{\Pi}}~\{-\hat{V}_K^{\sf DLPT}(\pi)\}=\inf_{\pi\in{\Pi}}~\{-\bbE_{S_2}[\psi(\bm z_i,\pi;\hat{\bm \theta}, \bm \Lambda)]\},
\end{equation}
where $\bbE_{S_2}[\cdot]$ denotes the empirical average over sample $S_2$.
The below result provides the regret guarantee of the learned policy. 

\begin{theorem}
\label{thm:learning_1}
 Define the function class: $\psi\circ \Pi = \{\psi(\cdot, \pi;\hat{\bm \theta}, {\bm \Lambda}):\pi\in\Pi\}.$
    Let $d$ be the VC subgraph dimension (also known as pseudo dimension) of  $\psi\circ \Pi$.     Let $r=\sup_{\pi\in\Pi}\|\psi(\cdot, \pi;\hat{\bm \theta}, {\bm \Lambda}) - \psi(\cdot, \pi^*;\bm \hat{\bm \theta}, {\bm \Lambda}) \|_{L_2(\bm X)}$.    Assume Assumptions \ref{ass:loss function}, \ref{ass:data bounded and nuisance smoonth}, and \ref{ass:influence function} in Appendix \ref{app:assumptions} hold, then with probability $1-\delta$,
      \begin{equation}
        R_K(\hat{\pi}^{\Pi}_{DLPT} ) = O\bigg(
         r\left(1+\sqrt{\log(1/r)}\right)\sqrt{\frac{d}{n}}+ r\sqrt{\frac{\log(1/\delta)}{n}}
        \bigg).
    \end{equation}
\end{theorem}

Theorem \ref{thm:learning_1} shows that the learned policy achieves minimax optimality, with regret scaling at the rate $1/\sqrt{n}$. This rate is consistent with results in policy learning literature for discrete treatment \citep{athey2021policy,zhou2023offline}. In other words, our model permits far more flexible functional forms and continuous optimization variables, yet its policy‑learning regret remains as efficient as in the original discrete‑experiment setting. The proof of Theorem \ref{thm:learning_1} is provided in Appendix \ref{appendix:regret bound proof}.

\begin{remark}[Role of policy class]
    The regret bound in Theorem \ref{thm:learning_1} explicitly accounts for policy class complexity, measured by VC subgraph dimension. This quantity captures a fundamental trade-off in policy learning: more complex policies are harder to learn effectively. However, characterizing the VC subgraph dimension can be challenging—it may even be infinite for some parametric function classes with finite-dimensional parameters \citep{shalev2014understanding}.
As an alternative, Appendix \ref{appendix:regret bound proof} presents a slightly more involved regret bound in Proposition \ref{prop:learning_2}, expressed in terms of covering numbers. Covering number is another standard measure of function class complexity and can be upper bounded by the VC subgraph dimension \citep{van2000asymptotic}. Note that covering numbers are often more tractable to analyze and have been studied for many common function classes, including those satisfying Lipschitz conditions \citep{vershynin2010introduction, xu2020upper}.
\end{remark}

In particular, we present
four distinct policy classes, spanning from industry standard to the most flexible option. All of them shall be evaluated in our empirical application:
\begin{itemize}
    \item \textit{\textbf{Finite Observed Policy Class:}} This class comprises a small, fixed number of  observed discrete treatment levels that have been experimented by the platform. Considering this class simplies our policy learning problem to discrete treatment scenarios, which have been carefully studied in literature \citep{athey2021policy,zhou2023offline}. The discrete nature reduces computational demands, making this class widely adopted in industry. However, it may prevent achieving higher potential outcomes that a more nuanced or continuous policy might achieve.
    \item \textit{\textbf{Discretized Continuous Policy Class:}} This class involves a discretization of a continuous range into evenly spaced intervals, introducing more granularity than the above  class with only experimented levels while still maintaining manageable complexity. 
    \item \textit{\textbf{Linear Policy Class:}} A class of functions defined by $t = \bm \alpha' \bm x + \beta$, where \( \bm{x} \) denotes the user's covariates, \( \bm{\alpha}  \) is a vector of policy coefficients that determine the weight assigned to each covariate, and \( \beta  \) is a scalar intercept term. 
    This class provides a straightforward way to scale decision-making based on measurable covariates, potentially aligning well with systematic variations in the data. The downside of this class is that the simplicity of linear assumptions may overlook complex nonlinear interactions within the data, potentially leading to suboptimal decisions.
     \item \textit{\textbf{DNN Policy Class:}} A class where $t = f(\bm x)$, with $f$ defined by a DNN. It offers the most flexibility and is capable of modeling complex and nonlinear relationships without predefined restrictions on the form of the policy function. This class can adaptively learn from data, offering the potential for high policy values by capturing intricate patterns that other classes may miss. However, the complexity and opacity of DNNs can lead to significant computational costs and may result in higher variability and uncertainty in policy outcomes, often requiring extensive data to train effectively.
     
\end{itemize}

\subsection{Model Misspecification and Corresponding Regret}
\label{sec:approximation_power}
Note that our method relies on identifying the optimal continuous policy within a flexible (semi-parametric) class of DGPs. However, despite the flexibility of our DGP specification in Equation~\eqref{eq:DGP}, misspecification remains possible. Therefore, in this section, we evaluate the approximation capability of the sigmoid-polynomial function assumed in Equation~\eqref{eq:DGP} relative to any DGP satisfying $s_t$-H\"older continuity (see Assumption \ref{ass:true_dgp_smooth} in Appendix \ref{app:assumptions}) defined as $f(\bm{x},t) := \mathbb{E}[Y \mid \bm{X} = \bm{x},\, T = t] $.



Assumption \ref{ass:true_dgp_smooth} imposes a smoothness condition on the ground-truth DGP function, which requires that $f(\bm x,t)$ is sufficiently smooth in the treatment variable $t$ uniformly over the context space $\mathcal{X}$. 
Under this condition, Theorem \ref{prop:representation power} establishes the approximation properties of the sigmoid-polynomial function $G(\bm{\theta}_K^*(\bm{x}),t)$, yielding the key result that the sigmoid-polynomial form can approximate the oracle DGP with a provable finite-sample error bound.
The proof of Theorem \ref{prop:representation power} is provided in Appendix \ref{appendix:prop approximation proof}.
\begin{theorem}[\sc Uniform Approximation Error]\label{prop:representation power}
    Suppose that the ground-truth DGP $f$ is $s_t$-H\"older continuous as described in Assumption \ref{ass:true_dgp_smooth}. Let $p$ be the degree of the approximation polynomial in $G(\bm\theta_K^*(\bmx),t)$.
    There exists coefficient function $\bm \theta^*_{K} (\bm x)\in \mathbb{R}^{K+1}$ such that
\begin{gather*}
    \sup_{(x,t)\in\mathcal{X}\times\mathcal{T}}\bigg|f(\bm x,t)-G(\bm\theta_K^*(\bmx),t)\bigg|=O(K^{-s_t}).
\end{gather*} 
\end{theorem}

 The polynomial component captures the variation of the function with respect to $t$, and under smoothness assumptions on the ground-truth DGP, standard approximation theory guarantees that a polynomial of degree $K$ can approximate the DGP with an error rate of $O(K^{-s_t})$. 

We conclude this section by emphasizing the trade-off involved in selecting the approximation degree $K$ and experiment implementation costs. A large $K$ can reduce the bias in value approximation, but its choice is constrained by identifiability, $K$ must not exceed the number of discrete treatment levels used in the experimental design (see Remark \ref{remark:selectK}). In practice, the number of treatment levels cannot be made arbitrarily large--especially in high-dimensional covariate settings--because this requires a sufficiently large sample to ensure adequate coverage of each treatment level across the covariate space. Otherwise, the estimation variance of the nuisance parameters increases, leading to higher optimization error and degraded policy performance. Thus, selecting $K$ involves balancing model complexity with available data.

\section{Empirical Setting, Experiment, and Main Analysis}\label{sec:empirical}
In this section, we introduce our collaborating platform, describe the large-scale randomized field experiment we conducted on the platform in order to evaluate our methodologies, and present the model-free empirical analysis of the experiment.


\subsection{Empirical Setting and Data}\label{sec:empirical_setting}

To empirically validate our proposed {\sf DLPT} framework, we partnered with Platform O, a leading global short-video platform. On this platform, users act as both content creators and viewers, with most videos being user-generated. Creators independently decide when and what content to produce, while viewers can access videos for free. Platform O primarily earns revenue through advertisements supplied by third-party advertisers. Creators do not receive a direct share of this ad revenue; instead, prominent creators monetize their content by selling products or securing direct advertising deals as influencers.

While larger creators can independently secure brand deals, smaller creators—especially newcomers—often lack the resources to do so. Yet, their participation is crucial for the platform's long-term success. To incentivize these creators, the platform has introduced targeted financial incentive programs. Platform O uses a proprietary digital token called “Points,” which users can earn through video gifts, monetization programs, and other engagement-based initiatives. These Points can be redeemed for cash-equivalent rewards. To incentivize small or new creators to upload videos, Platform O typically uses the ``Points Earning Task” feature to offer financial incentives, paid in points, for each uploaded video. This feature serves as the basis for our experimental intervention. 


\begin{figure}
    \centering
    \includegraphics[width=0.99\textwidth]{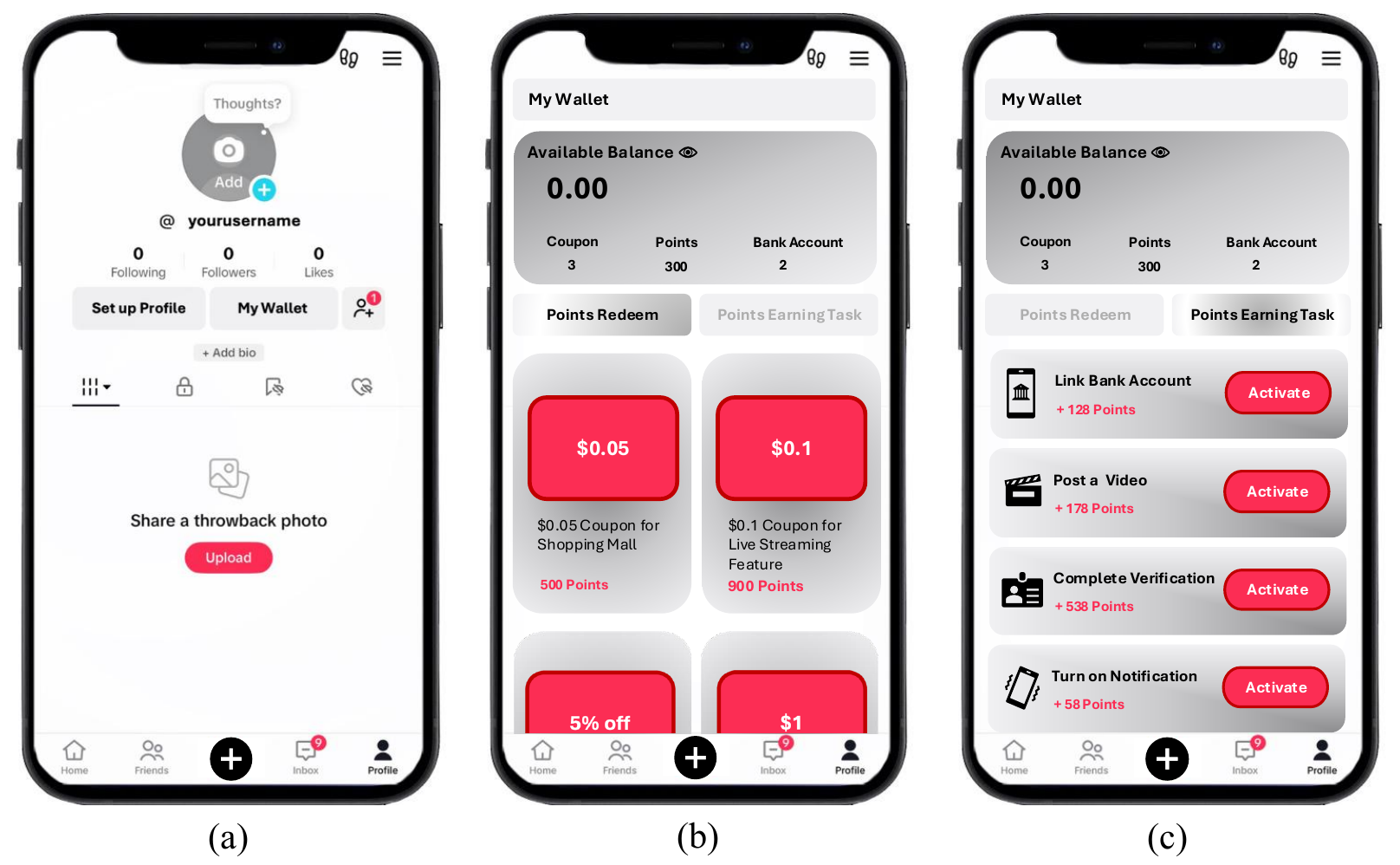}
    \caption{Illustration of the `Point" System on Platform}
    \subcaption*{\scriptsize Note: To protect Platform O’s identity and obscure nonessential details, we have slightly modified the interface of a widely used video-sharing platform.}
    \label{fig:task_center}
\end{figure}



\subsubsection{Experiment Design.} 

To examine how user-generated content production responds to financial incentives (specifically, reward Points) and to identify the optimal reward level for maximizing the platform’s return on investment, we conducted a randomized experiment from August 23 to 29, 2021, testing four distinct reward levels. The experiment targeted creators classified as low-activity producers, defined as users who had produced videos on fewer than three days during the four weeks preceding the experiment. These low-activity producers accounted for $12.3\%$ of all producers on the platform. All low-activity producers participated in the experiment and were randomly assigned to either a control group or one of the four treatment groups.

Users in the four treatment groups were offered a reward task that could be completed for varying levels of incentives during the experiment. This task, titled ``Post a Video," allowed users to earn a reward by posting at least one video during the experiment period (August 23 to 29, 2021). Users who posted multiple videos could earn the reward only once during the experiment. The reward levels randomly assigned to the four treatment groups were 178, 358, 538, or 718 Points, with each 100 Points approximately equivalent to \$0.01 USD. Users in the control group did not see this task and thus could not complete it. The specific user interface for this experiment is shown in Figure~\ref{fig:task_center}. As illustrated, users accessed their profile page (Panel (a)) and clicked on ``My Wallet" (Panel (b)) to view available Points Earning Tasks. Treatment group users saw the ``Post a Video" task with different Point values depending on their assigned group (Panel (c)), while control group users did not see this task at all. 

This feature potentially involves a trade-off: offering higher rewards increases the program’s cost but could also lead to more video production. The platform’s primary objective in conducting this experiment was to identify the optimal cost-effective level of incentives to encourage content production. While the incentive (i.e., Points awarded for video uploads) is conceptually a continuous variable, the platform could only implement and test a discrete set--in this case four levels--of treatments. These features of the experiment makes it a perfect test bed for our {\sf DLPT} framework. Last but not least, based on the experimental results, the platform selected the optimal discrete level and subsequently rolled out the feature to all users.

\subsubsection{Randomization Check and Data.} 

Due to data confidentiality, we only have access to a subsample of $7,349,648$ users in this experiment. Let  $t$  denote the treatment levels: the control group ($t$ = 0) included $1,478,472$ users, the 178 Points group ($t$ = 178) included $1,478,666$ users, the 358 Points group ($t$ = 358) included $1,479,277$ users,  the 538 Points group ($t$ = 538) included $1,478,819$ users, and the 718 Points group ($t$ = 718) included $1,479,414$ users. In our primary analyses,  each user $i$ is associated with an observed outcome $Y_i$, indicating whether the user posted at least one  video during the experimental period (August 23 to 29, 2021). Specifically, $Y_i = 1$ if the user posted at least one video, and $Y_i = 0$  otherwise.\footnote{Technically, $Y_i$ is a discrete variable taking values greater than or equal to zero. However, after discussions with the platform, it became clear that they are more concerned with whether a user publishes at least one video, rather than the exact number of videos posted. This preference is motivated by two reasons. First, only $9.53\%$ of low-activity users post more than one video within a week. Second, the platform's goal is to encourage users to develop a habit of content creation, for which posting a single video per week is deemed sufficient. Based on these considerations, and in alignment with the platform’s priorities, we follow their recommendation and define the outcome variable as binary.}  

For each user, we also collected a set of pre-treatment covariates, denoted  $\bm{X_i}$, which include 12 discrete variables (covering aspects such as gender, the user’s primary region of activity, the presence of other video-sharing platforms on the user’s device, and the user’s activity level) and 46 continuous variables (e.g.,  the number of videos uploaded by the user in the 30 days preceding the treatment). After applying one-hot encoding to the categorical variables, the dimensionality of the user pre-treatment covariates used in our analysis expanded to 77. Table \ref{tab:usercovariates} in the Appendix \ref{app:user_cov} provides a comprehensive description of all the covariates incorporated into our analysis.

To ensure the validity of the randomization process, we compared user demographics, content consumption, and content creation behavior during the 10 days prior to the experiment across all groups. As shown in Table~\ref{tab:random_check}, the five groups exhibited similar distributions in key demographic variables, such as gender and city tier. Furthermore, content consumption (measured by app usage) and content creation (measured by whether users uploaded at least one video) were also comparable across groups during the pre-experiment period. These results confirm that the randomization was successfully implemented, as there were no statistically significant differences in these characteristics across groups (all p-values $> 0.05$).
\begin{table}[t!]
    \centering\scriptsize
    \caption{Randomization Check}
\begin{tabular}{@{\extracolsep{2.5pt}} lcccccc} 
\\[-1.8ex]\hline 
\hline \\[-1.8ex] 

&&  {0} & {178} &  {358} & {538}  &  {718} \\
\hline\\[-1.8ex] 

\multirow{6}{3.5cm}{\textit{User Demographics}} & Proportion of  &42.38\%  & 42.39\% & 42.48\% & 42.31\% & 42.41\% \\
& Female Users &{} &(0.95) & (0.08) &(0.20) &(0.52)\\
& Proportion of &29.36\% &29.40\%&29.38\% & 29.38\% & 29.42\%  \\
& High-Active Users & &(0.49) &(0.76) &(0.75) &(0.23)  \\
& Proportion of Users &12.20\% &12.19\% &12.24\% &12.24\%&12.22\%\\
& from Moderately Developed City & &(0.85) &(0.25) &(0.22)&(0.44) \\

\hline \\[-1.8ex] 
\multirow{4}{3.5cm}{\textit{User Behaviors 10 Days Prior to the Experiment}} 
& Average App Usage & 0.0010&-0.0000&0.0005&0.0011&-0.0027\\
&Duration per User&&(0.34) &(0.67) &(0.93)&(0.95) \\
& Video Uploading or Not& -0.0014&0.0003&0.0005&-0.0007&0.0002\\
&&&(0.15) &(0.11) &(0.52)&(0.11) \\

\hline
\hline \\[-1.8ex] 

\end{tabular} 
\begin{tablenotes}
\item Note: p-values of t-tests between the control group $t$ = 0 and other levels of points are reported in parentheses. The reported metrics of app-usage duration and at least one video uploading or not are rescaled to protect sensitive data.
\end{tablenotes}
    \label{tab:random_check}
\end{table}
\subsection{Model-Free Empirical Evidence}\label{sec:empirical_effects}
For this experiment to serve as a suitable testbed for our {\sf DLPT} framework, it must satisfy two key conditions beyond its experimental design. First, the treatment—in this case, the financial reward—must have a statistically significant impact on the outcome variable, i.e., whether to upload a video during the experiment. Second, the effect of the treatment should vary across different reward levels and user subgroups, which creates the room for personalization. Accordingly, in this subsection, we present the average treatment effects (ATEs) and heterogeneous treatment effects (HTEs) from the original experiment prior to implementing our framework.

\subsubsection{The Average Impact of Financial Reward on Content Production.}
We first provide empirical evidence that examines the ATEs of receiving Points on the recipient’s content production. For each of the four treated Point levels, we compared user outcomes against the control group ($t = 0$) to calculate the ATE for each level. 
We use the following ordinary least squares regression specification:
\begin{equation}\label{eq:ate_regr}
    Y_i = \alpha_0+\alpha_1 D_i + \epsilon_i,
\end{equation}
where $Y_i$ is the binary outcome variable indicating whether user $i$ posted at least one public video during the experiment week, and $D_i$ is the binary treatment variable indicating whether user $i$ is assigned to the treated group. For each tested Point level, we estimate  specification \eqref{eq:ate_regr} using data from control and treated users at that specific Point level.

\begin{table}[!t]\centering \scriptsize
\caption{Average Treatment Effects of Point Rewards}
\setlength{\tabcolsep}{4.0mm}{
\begin{tabular}{cccccc}
    \hline 
    \hline \\[-1.8ex] 
 & \multicolumn{4}{c}{Dependent Variable: Video Uploading or Not}      & Max Pairwise $p$-value       \\
 \cmidrule{2-5}
&   178     & 358     & 538     & 718 & \\  \hline  \\[-1.8ex] 
Treatment & 0.0586$^{****}$ & 0.0729$^{****}$ & 0.0824$^{****}$ &0.0886$^{****}$ & $<1.189e^{-07}$\\
&(0.00113)&(0.00114)&(0.00115)&(0.00115)&\\
\hline\\[-1.8ex] 
Relative Effect Size   & 13.01\% & 16.20\% & 18.30\% & 19.69\%&\\
Observations & 2,957,138 & 2,957,749& 2,957,291 &2,957,886 &\\
    \hline 
    \hline  \\[-1.8ex]   
\end{tabular}}\label{tab:ground-truth ate}
\begin{tablenotes}
\item Note: To preserve data confidentiality, we normalize the dependent variable in this table. Note that normalization is applied only to present the empirical results. The original binary outcome is retained when implementing the {\sf DLPT} framework. We report the relative effect size of each reward level, scaled by the mean outcome in the control group. Robust standard errors are reported in parentheses. $^{*}p<0.05$;$^{**}p<0.01$;$^{***}p<0.001$;$^{****}p<0.0001$.
\end{tablenotes}
\end{table}

As shown in Table~\ref{tab:ground-truth ate}, the probability of users posting a video during the experimental period was significantly higher in all four treatment groups compared to the control group (all $p$-values $< 0.0001$), with relative effect sizes ranging from $13.01\%$ to $19.69\%$. Moreover, the upload probability increased with higher reward amounts, and significant differences were observed across different reward levels. In the last column of Table~\ref{tab:ground-truth ate}, we report the maximum pairwise $p$-value from comparisons between adjacent reward levels (e.g., 178 vs. 358, 358 vs. 538, and 538 vs. 718); this  $p$-value is $< 1.189 \times 10^{-7}$, indicating that the treatment effects significantly vary across reward levels and that larger rewards lead to greater increases in upload probability. Consistent with prior literature~\citep{kaynar2023estimating}, we also observe diminishing marginal returns to financial incentives in promoting content creation: although the incentive roughly doubles at each level, the marginal increase in user participation decreases from 3.19\% to 2.10\%, and eventually to 1.39\%.


\subsubsection{The Heterogeneous Impact of Financial Reward on Content Production.}\label{sec:HTE}
We further investigate the HTEs of reward levels across user subgroups. Specifically, we identify two user characteristics that may moderate the effect of Points rewards on content production. First, users’ historical video uploading behavior may influence their responsiveness to reward incentives. Second, inspired by \cite{zeng2024impact}, we consider users’ content consumption behavior, which may also affect their likelihood of content production. To examine the two potential moderators, we construct two binary variables. The first, $LowP_i$, captures historical content production behavior and equals one if user $i$ uploaded fewer videos in the 30 days prior to the experiment than the median user in the sample. The second, $LowC_i$, reflects historical content consumption behavior and equals one if user $i$ completed watching fewer videos in the 30 days prior to the experiment than the sample median.

We rely on the following OLS regression specifications to test the interaction between the treated Point rewards and two potential moderators, using data from control and treated users at that specific Point level:
\begin{gather}\label{eq:hte_regr}
      Y_i = \beta_0+\beta_1 D_i+ \beta_2 D_i\times LowP_i+\beta_3 LowP_i +\epsilon_i\\
      Y_i = \gamma_0+\gamma_1 D_i+ \gamma_2 D_i\times LowC_i+\gamma_3 LowC_i +\epsilon_i
\end{gather}


For data confidentiality, we normalize the outcome variable. Table~\ref{tab:HTE_effect} presents the HTEs of different Point reward levels on users' content production. Across all reward levels, we observe significant positive treatment effects, with effect sizes increasing as the reward amount rises. Importantly, users' historical production and consumption behaviors significantly moderate these effects. Specifically, users classified as low producers ($LowP$) exhibit a consistently lower responsiveness to rewards compared to high producers, with negative and statistically significant interaction terms across all reward levels (all $p$-values$<0.0001$). Conversely, users classified as low consumers ($LowC$) demonstrate a significantly higher responsiveness to rewards, as indicated by positive and significant interaction terms (all $p$-values$<0.01$). These patterns suggest that user heterogeneity plays a meaningful role in shaping the effectiveness of reward interventions. Consequently, these results provide strong motivation for personalized reward targeting strategies.

\begin{table}[!t]
    \centering\scriptsize
    \caption{Heterogeneous Treatment Effects of Point Rewards}
    \setlength{\tabcolsep}{1.5mm}{
    \begin{tabular}{lcccccccc}
\hline
\hline\\[-1.8ex] 
& \multicolumn{8}{c}{Dependent Variable: Video Uploading or Not}\\
\cmidrule{2-9}
& \multicolumn{2}{c}{178} & \multicolumn{2}{c}{358} & \multicolumn{2}{c}{538}& \multicolumn{2}{c}{718}     \\
\cmidrule{2-9}
&(1a)&(1b)&(2a)&(2b)&(3a)&(3b)&(4a)&(4b)\\
\hline  \\[-1.8ex] 
Treatment &0.0684$^{****}$& 0.0529$^{****}$ & 0.0803$^{****}$ &0.0637$^{****}$ & 0.0889$^{****}$& 0.0763$^{****}$ &0.0919$^{****}$ & 0.0812$^{****}$  \\
&((0.00205)&(0.00176)&(0.00205)&(0.00176)&(0.00206)&(0.00177)&(0.00206)&(0.00177)\\
Treatment$\times$LowP&-0.0183$^{****}$&&-0.0128$^{****}$&&-0.0121$^{****}$&&-0.0068$^{**}$&\\
&(0.00233)&&(0.00235)&&(0.00235)&&(0.00236)&\\
Treatment$\times$LowC&&0.0112$^{****}$&&0.0185$^{****}$&&0.0126$^{****}$&&0.0150$^{****}$\\
&&(0.00225)&&(0.00226)&&(0.00227)&&(0.00227)\\
\hline\\[-1.8ex] 
Observations& \multicolumn{2}{c}{2,957,138} & \multicolumn{2}{c}{2,957,749}& \multicolumn{2}{c}{2,957,291} &\multicolumn{2}{c}{2,957,886} \\
   \hline 
    \hline  \\[-1.8ex]   
\end{tabular}}
\label{tab:HTE_effect}
\begin{tablenotes}
\item Note: To preserve data confidentiality, we normalize the dependent variable in this table. Note that normalization is applied only to present the empirical results. The original binary outcome is retained when implementing the {\sf DLPT} framework. Robust standard errors are reported in parentheses. $^{*}p<0.05$;$^{**}p<0.01$;$^{***}p<0.001$;$^{****}p<0.0001$.
\end{tablenotes}
\end{table}

\section{Evaluation with the Field Data}\label{sec:eval_empirical}
In this section, we empirically evaluate the {\sf DLPT} framework using the field experiment data described in Section~\ref{sec:empirical}. Section~\ref{sec:empirical_application} outlines the implementation details and benchmarks. Section~\ref{sec:ATE_recover} presents ATE estimation results for the four tested Point levels. Section~\ref{sec:disc_policy} reports policy learning regret for the evaluation user groups under the discrete policy class. 


\subsection{Framework Implementation and Benchmark}\label{sec:empirical_application}
To implement the {\sf DLPT} framework in our empirical setting, we follow the three-stage procedure outlined in Section~\ref{sec:theo_framework}. We denote the dataset of $7,349,648$ users as $\mathcal{S}$ and partition it into a training set $\mathcal{S}_{\text{train}}$ and an inference set $\mathcal{S}_{\text{inference}}$. The specific splitting strategy is designed to align with our evaluation objectives, which we will elaborate on in the following sections. Nevertheless, the implementation procedure itself is general and remains applicable across different evaluation setups.

\begin{figure}
    \centering
    \includegraphics[width=0.99\textwidth]{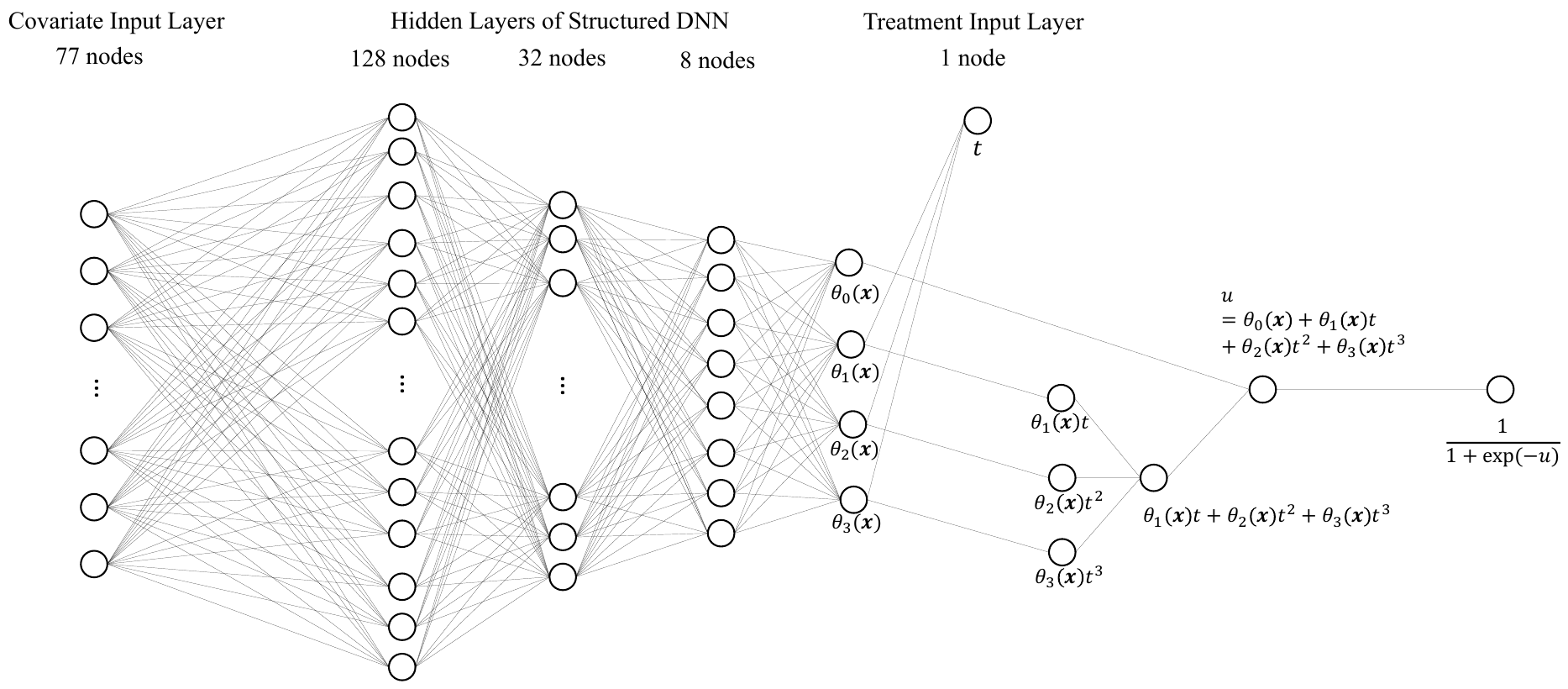}
    \caption{Illustration of Structured Deep Neural Network in Empirical Application}
    \label{fig:dnn_structure}
\end{figure}

\textbf{Stage 1: Nuisance Parameters Estimation}. In Stage 1, we assume that the outcome $Y$, user features $\bm{X}$, and continuous treatment level $T$ (rescaled by dividing by 1000 for  training stability) are generated according to the DGP specified in Equation (\ref{eq:DGP_ass}), with the polynomial degree set to $K=3$:
\begin{equation}\label{eq:G_func_3}
    \mathbb{E}[Y | \bm X = \bm x, T= t]=G(\bm{\theta}^*(\bm{x}),t) = \frac{1}{1+\exp(-(\theta^*_0(\bm x)+\theta^*_1(\bm x)t +\theta^*_2(\bm x)t^2+ \theta^*_3(\bm x)t^3))}.
\end{equation}
The policy value for a given policy $\pi$ is specified as follows:
\begin{equation*}\label{eq:policy_value_empirical}
    V(\pi)=\bbE[wY-c\pi(\bm X)],
\end{equation*}
where $w=0.5$ and $c=0.1$ are parameters set by Platform O. 
These represent the per-unit revenue (in Point) from users uploading videos and the per-unit cost (also in Point) of providing monetary incentives, respectively. 
Thus, $V(\pi)$ captures the net value, as the revenue from user activity minus the cost of incentives, measured entirely in Points.
Figure \ref{fig:dnn_structure} illustrates the structured DNN that approximates the DGP in Equation \eqref{eq:G_func_3}, implemented via TensorFlow. The training dataset $\mathcal{S}_{\text{train}}$ is used to fit the structured DNN, which is trained under a binary cross-entropy loss function through the ADAM optimizer \citep{kingma2014adam}.


\textbf{Stage 2: Policy Value Estimation.} Next, we construct estimators for two quantities: (i) the ATE of treatment level $t$ relative to the control, denoted as $\mu(t)$; and (ii) the policy value of a given policy $\pi$, denoted  $V_K(\pi)$. We estimate the policy value for policy learning in the next stage, whereas we estimate the ATE of observed treatment primarily for evaluation purposes.
\begin{gather*}
    \mu(t)= \bbE[H_{\mu}(\bm\theta(\bmx),t)]=\bbE[G(\bm\theta(\bm x),t)-G(\bm\theta(\bm x),0)],\\
    V_K(\pi)=\bbE[H_{V}(\bm\theta(\bm x),\pi(\bm x))]=\bbE[wG(\bm\theta(\bm x),\pi(\bm x))-c\pi(\bm x)]
\end{gather*} 
 We then construct Neyman Orthogonal score function for the ATE estimator using testing data $\mathcal{S}_{\text{inference}}$ following Proposition \ref{prop:influence function} (since the non-parametric representation in Stage 1 is trained on $\mathcal{S}_{\text{train}}$):
\begin{gather*}
   \hat{\mu}(t) = \frac{1}{|\mathcal{S}_{\text{inference}}|}\sum_{i \in \mathcal{S}_{\text{inference}}} \psi_\mu (\bm z_i,t; \hat{\bm\theta},\bm \Lambda) \\
    \text{where } \psi_\mu (\bm z_i,t; \hat{\bm\theta},\bm \Lambda) = H_\mu(\hat{\bm\theta}(\bm x_i),t) - H_{\mu{\bm\theta}}(\hat{\bm\theta}(\bm x_i), t)\bm \Lambda(\bmx_i)^{-1} \ell_{\bm\theta}(y,\check{t}_i, \hat{\bm\theta}(\bm x_i)).
\end{gather*}
Similarly, the Neyman Orthogonal score function for the policy value estimator is given by:
\begin{gather*}
   \hat{V}_K^{\sf DLPT}(\pi) = \frac{1}{|\mathcal{S}_{\text{inference}}|}\sum_{i \in \mathcal{S}_{\text{inference}}} \psi_V (\bm z_i,\pi; \hat{\bm\theta},\bm \Lambda) \\
   \text{where } \psi_V (\bm z_i,\pi; \hat{\bm\theta},\bm \Lambda) = H_V(\hat{\bm\theta}(\bm x_i),\pi(\bm x_i)) - H_{V\bm\theta}(\hat{\bm\theta}(\bm x_i), \pi(\bm x_i))\bm\Lambda(\bm x_i)^{-1} \ell_{\bm\theta}(y,  \check{t}_i, \hat{\bm\theta}(\bm x_i)).
\end{gather*} Here $\check{t}_i$ denotes the treated level of points of the evaluation data. 
\textbf{Stage 3: Policy Learning.}
We derive the personalized optimal policies by solving the ERM problem $ \hat{\pi}_K^{DLPT} = \inf_{\pi\in{\Pi}}~\{-\hat{V}_K^{\sf DLPT}(\pi)\}$ for inference data $\mathcal{S}_{\text{inference}}$ under four different policy classes $\Pi$: 
\begin{enumerate}
    \item Finite Observed Policy Class: This class includes five observed discrete levels of intervention Points $\{0,178,358,538,718\}$
    \item Discretized Continuous Policy Class: This class discretizes the continuous range $[0,1000]$ into 100 Points, enumerated as $\{0, 10,20, \dots, 990, 1000\}$.
    \item Linear Policy Class: A class of functions defined as $t = \bm \alpha' \bm x + \beta$, where $\bm x$ represents users' covariates.
    \item DNN Policy Class: A class of functions where $t = f(\bm x)$, and $f$ is defined by a DNN. Specifically, the DNN is a three-layer, fully connected network with 10 nodes per layer.
\end{enumerate} 

For the first two discrete policy classes, we compute the value of $ \psi_V (\bm z_i,\pi; \hat{\bm\theta},\bm \Lambda)$ for each candidate policy $\pi \in \Pi$, given user covariates $\bm x_i$, and assign to user $i$ the policy that yields the highest estimated value. For the last two policy classes, we optimize the objective $\hat{V}_K^{\sf DLPT}(\pi)$ directly using TensorFlow's automatic differentiation \citep{baydin2018automatic}, treating it as the loss function to train the policy classes.
See Appendix \ref{app:empirical_implementation} for implementation details. 

\textbf{Benchmarks.}
To evaluate the effectiveness of our framework in estimating the ATE and, more importantly, in enabling personalized policy learning, we compare it against five established benchmarks: (a) linear regression ({\sf LR}); (b) logistic regression ({\sf LogR}); (c) the logit choice model ({\sf LCM}); (d) a pure deep learning approach ({\sf PDL}); and (e) a uniform level policy ({\sf UNI}).

Under the {\sf LR} approach, it assumes a linear relationship between the expected outcome and the covariates and treatment such that $\bbE[Y|\bm X=\bmx, T = t]=\bm\alpha\bmx+\beta t$. This linear functional form is a common and interpretable baseline for estimating treatment effects in empirical settings, particularly in market intervention effectiveness and pricing studies (e.g., \citealp{lodish1995tv,besbes2015surprising,simester2020efficiently,cui2020reducing}).

When the outcome $Y$ is binary (uploading videos or not), {\sf LogR} is a more appropriate choice. {\sf LogR} models the probability of the outcome as a sigmoid function of a linear combination of covariates and treatment: $\bbE[Y|\bm X=\bmx, T = t]=\sigma(\bm\alpha\bmx+\beta t)$, where $\sigma(x)=\frac{1}{1+\exp(-x)}$.  We remark that while {\sf LR} inherently assumes the linear boosting effect of point policy, {\sf LogR} accounts for the pattern of diminishing returns. This is widely used for modeling customer response to interventions (e.g., \citealp{guadagni1983logit,mela1997long,bleier2015personalized,ascarza2017beyond}).

The \textsf{LCM} approach assumes that the utility a user derives from uploading videos, given their features $\bm x$ and point level $t$, is $u = \bm\alpha \bmx +\bm\beta \bm x t+\gamma$, where $\gamma$ represents a Type-I extreme value error term. This leads to the probability of uploading as $\bbP[Y=1|\bm X=\bm x,T=t]=\frac{\exp(u)}{1+\exp(u)}=\sigma(u)$. This model, rooted in discrete choice theory, is a widely adopted benchmark for understanding customer behavior (e.g., \citealp{tian2020optimizing,dube2023personalized,el2023joint}). While \textsf{LogR} and \textsf{LCM} share the same underlying logistic functional form, a crucial distinction is \textsf{LCM}'s ability to model a heterogeneous effect of treatment on the outcome through the interaction term $\bm\beta \bm x t$.

The {\sf PDL} use the similar DNN as shown in Figure \ref{fig:dnn_structure} to approximate $\bbE[Y|\bm X=\bmx, T = t]=f_{\sf PDL}(\bmx,t)$. Unlike the parametric assumptions of the previous models, PDL offers a non-parametric approach to capture complex, non-linear relationships between inputs and outcomes. With the increasing abundance of data, the deep learning approach has been widely adopted (e.g., \citealp{simester2020targeting,garcia2022demand,yoganarasimhan2023design}). 

The {\sf UNI} policy, as the current industry practice, assigns a uniform treatment level to all individuals, typically selected by maximizing the estimated policy value derived from the ATE calculated across the observed sample.

The first four methods---{\sf LR}, {\sf LogR}, {\sf LCM}, and {\sf PDL}---are used for both ATE estimation and policy learning. The final benchmark, {\sf UNI}, is limited to policy learning and assigns the same treatment level to all individuals. We include {\sf UNI} as a baseline to assess the added value of personalized, data-driven policy optimization.
The implementations of all benchmarks use the same training dataset $\mathcal{S}_{\text{train}}$ and inference dataset $\mathcal{S}_{\text{inference}}$. See Appendix \ref{app:imple_bench_real} for detailed benchmark descriptions and implementations.

\subsection{Results on ATE Evaluation}\label{sec:ATE_recover}
In this section, we present the ATE estimation result in Stage 2, which  serves as an indicator of the accuracy of the estimated nuisance parameters in Stage 1.
 Specifically, we adopt a cross-evaluation strategy: for each Point Reward level $t \in \{178, 358, 538, 718\}$, we designate users assigned to treatment level $t$ as the masked evaluation set $\mathcal{S}_{\text{masked}}$, which outcomes are used for computing the ground truth  ATE of the masked level. 
 The remaining users are then randomly split into a training set $\mathcal{S}_{\text{train}}$ (for estimating nuisances in Stage 1) and an inference set $\mathcal{S}_{\text{inference}}$ (for constructing orthogonalized ATE estimator in Stage 2). 
 The ATE estimation results for each level $t$ are reported in Table~\ref{tab:realdata_ate_result}.  Column (2) presents the ground-truth normalized ATE for the four levels of Points. Columns (3)–(12) report the estimated ATEs and their corresponding absolute percentage errors (APE) against the ground-truth ATEs shown in Column (2). To evaluate the accuracy of the estimates, we use three performance metrics to aggregate the performance of different methods across the $4$ test cases: Mean Absolute Percentage Error (MAPE), Mean Squared Error (MSE), and Mean Absolute Error (MAE).

\begin{table}[!t]
\centering \scriptsize
\caption{Comparison of ATE Estimators Across Methods}
\setlength{\tabcolsep}{1.3mm}{
\begin{tabular}{cccccccccccc}
\\[-1.8ex]\hline 
\hline \\[-1.8ex] 
&& \multicolumn{2}{c}{\sf LR}      & \multicolumn{2}{c}{\sf LogR}    & \multicolumn{2}{c}{\sf LCM}     & \multicolumn{2}{c}{\sf PDL}       & \multicolumn{2}{c}{\sf DLPT}  \\
\cmidrule{3-12}
Level of Points & Ground-Truth ATE & Est.&APE & Est.&APE  & Est.&APE  & Est.&APE & Est.&APE \\  
(1)&(2)&(3)&(4)&(5)&(6)&(7)&(8)&(9)&(10)&(11)&(12)\\[-1.8ex] \\
\hline  \\[-1.8ex] 
178  & 0.0586$^{****}$ & 0.0225  & 61.63\%  & 0.0214  & 63.51\% & 0.0214 & 63.50\% &0.0492& 15.99\% & 0.0603 & 3.00\%\\
358  & 0.0729$^{****}$ & 0.0402  & 44.91\%  & 0.0446 &38.85\% & 0.0342& 53.04\% &0.0707&3.01\%&0.0719 & 1.40\%\\
538  & 0.0824$^{****}$  & 0.0596 & 27.73\% & 0.0533 & 35.37\% &  0.0595&27.80\%&0.0795 & 3.51\% &0.0813 & 1.32\% \\
718  & 0.0886$^{****}$  & 0.1051 & 18.66\% & 0.1035 & 16.77\% & 0.1152& 30.02\%&0.0831 &6.22\% &0.0857 & 3.30\% \\[-1.8ex] \\  \hline  \\[-1.8ex] 
\multicolumn{2}{c}{MAPE} & &38.23\% && 38.60\% && 43.59\% &&7.18\%  && 2.25\%  \\
\multicolumn{2}{c}{MSE}  & &792.99  && 814.35  &&1028.02  && 32.82  && 3.46   \\
\multicolumn{2}{c}{MAE}  & &27.06   &&27.39 &&31.35 &&  4.99 &&1.70 \\[-1.8ex] 
\\\hline 
\hline \\[-1.8ex] 
\end{tabular}}
\begin{tablenotes}
\item Note: To preserve data confidentiality, we normalize the dependent variable in this table in Column (2) as well as the estimators of all methods. Note that normalization is applied only to present the results. The original binary outcome is retained when implementing the {\sf DLPT} framework. MSE is scaled by multiplying a constant. MAE is scaled by multiplying another constant.
\end{tablenotes}
    \label{tab:realdata_ate_result}
\end{table}

As shown in Table \ref{tab:realdata_ate_result}, across all levels, {\sf DLPT} consistently achieves the lowest APEs, with values ranging from 1.32\% to 3.30\%, significantly outperforming baseline methods such as {\sf LR} (18.66\%–61.63\%) and {\sf LCM} (27.80\%–63.50\%). The summary metrics at the bottom of the table further highlight {\sf DLPT}'s superior accuracy, yielding the lowest MAPE (2.25\%), MSE (3.46), and MAE (1.70). Among baseline methods, {\sf PDL} also performs competitively, achieving 7.18\% in MAPE and outperforming traditional models such as {\sf LR}, {\sf LogR}, and {\sf LCM} by a large margin. These results demonstrate the effectiveness of {\sf DLPT} in accurately estimating treatment effects, which is critical for downstream policy learning. 

Unlike {\sf LR} and {\sf LogR}, which rely on linear assumptions or simple parametric transformations, {\sf DLPT} leverages the representational power of deep learning combined with bias correction techniques to achieve more accurate ATE estimates. While {\sf PDL} also benefits from the flexibility of neural networks and performs considerably better than parametric models, it does not incorporate explicit debiasing procedures. As a result, its estimation errors, though relatively low, remain consistently higher than those of {\sf DLPT}. This highlights the importance of correcting for estimation bias in high-capacity models when recovering treatment effects.

Finally, it is noteworthy that both {\sf DLPT} and {\sf PDL} exhibit slightly higher estimation errors at the boundary treatment levels ($t=178$ and $t=718$) compared to the intermediate levels ($t=358$ and $t=538$), a pattern not observed with more parametric methods. We interpret this as suggestive evidence that these two deep learning methods tend to overfit within the observed parameter range, implying that extrapolation for DL-based methods is inherently more challenging than interpolation. Consequently, when employing these methods in practice, experimenters should design initial experiments to span a sufficiently large parameter space, thereby mitigating the need for extrapolation.

\subsection{Results on Policy Learning under Discrete Policy Class} \label{sec:disc_policy}

We now present the results from Stage 3 on personalized policy learning, focusing on the discrete policy class, specifically, the Finite Observed Policy Class. To ensure minimal approximation error, we set the polynomial degree $K=4$ and change the structured DNN accordingly. We will extend the analysis to the four policy classes, including continuous policies, in Section~\ref{sec:real_data_synthetic}. We adopt a similar cross-evaluation logic as in Section~\ref{sec:ATE_recover} in the implementation but also add a random partitioning of the user subgroups into training and inference datasets. To enable this, we discretize user covariates and partition users into $2,287$ subgroups based on six demographic and behavioral characteristics: gender, developmental level of the user’s primary city of residence (categorized as highly developed, moderately developed, or less developed), user activity level (low, medium, or high), platform usage (iPhone or other), life stage (e.g., mature phase), and number of friends. Users within each subgroup share identical values across these dimensions. We then split all users into the training and inference datasets at the group level randomly. The resulting training and inference datasets are  approximately equal in size and randomly composed, denoted as $\mathcal{S}_{\text{train}}$ (with 1,112 subgroups containing 3,681,464 users) and $\mathcal{S}_{\text{inference}}$ (with 1,175 subgroups containing 3,668,184 users).

To evaluate our method against benchmark approaches, we assume that as experimenters we observe the outcome variables in $\mathcal{S}_{\text{train}}$ but not those in $\mathcal{S}_{\text{inference}}$. Consequently, to implement our proposed method, we first use $\mathcal{S}_{\text{train}}$ in Stage~1 to obtain $\hat{\theta}(\cdot)$ and then apply Stages~2 and~3 using features from $\mathcal{S}_{\text{inference}}$. For each subgroup \( j \) in $\mathcal{S}_{\text{inference}}$, we compute the estimated optimal policy—that is, the optimal point level among $\{0,178,358,538,718\}$—by selecting the point level that yields the highest estimated value according to our Stage~3 estimator. Similarly, for the benchmark methods, we use $\mathcal{S}_{\text{train}}$ to estimate parametric policy values based on features and then determine the optimal policy using features from $\mathcal{S}_{\text{inference}}$. Finally, for each subgroup \( j \) in $\mathcal{S}_{\text{inference}}$, we calculate the ``ground-truth'' policy value for each point level using empirical average outcomes, and we determine the ``ground-truth'' optimal policy as the point level that yields the highest average outcome within that subgroup. We then compare the values of each method's estimated optimal policy and the ground-truth optimal policy for each subgroup.

Table~\ref{tab:realdata_regret_result} compares the performance of the proposed {\sf DLPT} framework with all benchmark methods. Performance is evaluated based on aggregating results through all subgroups in $\mathcal{S}_{\text{inference}}$ using three metrics: (1) Mean Percentage Regret (MPR), which measures regret relative to the optimal in-class policy value; (2) accuracy, defined as the percentage of subgroups for which the learned policy correctly identifies the optimal Point level; and (3) group-size-weighted accuracy, which similarly compares the learned best Point level to the true best Point level, but weights each subgroup’s contribution by its user size.

As shown in Table~\ref{tab:realdata_regret_result}, the proposed {\sf DLPT} framework achieves the best performance by a significant margin. It obtains the lowest MPR of $0.48\%$, indicating that its obtained policy is only $0.48\%$ lower than the ground-truth optimal policy. {\sf DLPT} also substantially outperforms all baselines in both accuracy (73.08\%) and group-size-weighted accuracy (79.83\%), reflecting its ability to correctly identify the best policy for both small and large user subgroups. The {\sf UNI} method, which selects the optimal point level with the highest average outcome across all users (thus ignoring heterogeneity), performs moderately well with an MPR of 6.41\% and accuracy around 25--34\% (note that random guessing among five treatment levels yields an accuracy of 20\%). Surprisingly, traditional models such as {\sf LR}, {\sf LogR}, and {\sf LCM} perform worse than {\sf UNI} across all metrics. This suggests that the rigid functional forms of these traditional models may be overwhelmed by the large number of user features included. Even {\sf PDL}, a more flexible baseline, does not match the performance of {\sf DLPT}, particularly in accuracy. These results collectively underscore the effectiveness of our {\sf DLPT} framework in policy learning, highlighting the importance of modeling heterogeneity flexibly and correcting for biases in the estimation of heterogeneous treatment effects.

\begin{table}[!t]
\centering \scriptsize
\caption{Comparison of Policy Regret Across Methods}
\setlength{\tabcolsep}{5mm}{
\begin{tabular}{ccccccc}
\\[-1.8ex]\hline 
\hline \\[-1.8ex] 
 &{\sf LR}      & {\sf LogR}    & {\sf LCM}     & {\sf PDL}   &{\sf UNI}   &{\sf DLPT}  \\  
&(1)&(2)&(3)&(4)&(5)&(6)\\[-1.8ex] \\
\hline  \\[-1.8ex] 
MPR&7.47\%  & 10.05\%    &9.91\%  & 6.32\%  & 6.41\%& 0.48\%   \\
Accuracy & 24.55\% &20.85\% &20.19\% & 25.50\%& 24.74\% &73.08\%  \\
Group-Size-Weighted Acc. &31.61\%&19.64\%&13.50\%&32.70\%& 33.79\%& 79.83\% \\
\hline 
\hline \\[-1.8ex] 
\end{tabular}}
    \label{tab:realdata_regret_result}
\end{table}

\section{Evaluation with Semi-Synthetic Data}\label{sec:semi-syn}
In this section, we extend our evaluation of the {\sf DLPT} framework through semi-synthetic studies grounded in real-world field experiment data. It allows us to evaluate policy learning performances when considering a broader range of policy classes beyond the experimental setting. In Section~\ref{sec:fnn_dgp_appr}, we describe the training of a fully connected deep neural network (FNN) on the complete dataset of $7,349,648$ users; this FNN serves as the  semi-synthetic DGP we use to approximate the ground-truth DGP. Section~\ref{sec:real_data_synthetic} presents the results on personalized policy learning under continuous policy classes.  Finally, Section~\ref{sec:robustness} provides a series of robustness checks to assess the performance of our framework.

\subsection{Ground-Truth DGP Approximation and Semi-Synthetic Data Generation}\label{sec:fnn_dgp_appr}
To approximate the ground-truth DGP, we train a FNN on the complete dataset of $7,349,648$ users such that $\bbE[Y|\bm X, T]=f^{\text{FNN}}(\bm X,T)$. Both user feature $\bm X$ and assigned treatment level $T$ are the input, and binary outcome $Y$ is the output. The FNN architecture consists of one input layer with 78 nodes (with $\bm X$ in 77 dimensions and $T$ in 1 dimension), two fully connected layers with $4,096$ nodes per layer, and one output layer with 1 node and sigmoid as activation function, resulting in a total of $17,108,993$ parameters. We use binary cross-entropy as the loss function, adopt the Nadam optimizer \citep{dozat2016incorporating}, and set the batch size to $1,024$. The model is trained for 100 epochs, achieving a final accuracy of 99.80\%. While this high-capacity configuration carries a risk of overfitting, it enables flexible simulation across the entire continuous treatment space and closely approximates the ground truth DGP for all users in the dataset, facilitating robust evaluation across various continuous policy classes.

It is important to note that the FNN does not need to perfectly recover the ground-truth DGP or attain 100\% predictive accuracy for the evaluation to be valid. The evaluation validity holds because we use the FNN both to generate training data for policy learning methods and to evaluate the learned policies based on the data from the FNN. However, achieving a high-fidelity approximation of the FNN is still desirable since the closer the FNN matches the ground-truth DGP, the more accurately we can assess the approximation power of the {\sf DLPT} framework in practice. This, in turn, increases our confidence in the framework’s performance under realistic misspecification in real-world deployment, as discussed in Section~\ref{sec:approximation_power}.

To generate semi-synthetic data $\mathcal{S}^{\text{FNN}}$, we use the trained $f^{\text{FNN}}(\cdot)$ to obtain the probability of outcome variable $Y$ given $\bm X$ and $T$. For all $7,349,648$ users, we generate $Y^{\text{FNN}}_i = B(1,f^{\text{FNN}}(\bm X_i,T_i))$ following the binomial distribution with $\mathbb{P}(Y^{\text{FNN}}_i=1)=f^{\text{FNN}}(\bm X_i,T_i)$. The $\mathcal{S}^{\text{FNN}}$ is randomly half split into  a training set \( \mathcal{S}^{\text{FNN}}_{\text{train}} \) and a testing set \( \mathcal{S}^{\text{FNN}}_{\text{testing}} \). We then randomly half-split testing set \( \mathcal{S}^{\text{FNN}}_{\text{testing}} \) into inference set \( \mathcal{S}^{\text{FNN}}_{\text{inference}} \) and evaluation set \( \mathcal{S}^{\text{FNN}}_{\text{evaluation}} \). We use the training data \( \mathcal{S}^{\text{FNN}}_{\text{train}} \) to for stage 1 estimation, inference data \( \mathcal{S}^{\text{FNN}}_{\text{inference}} \) for stages 2 and 3; we then evaluate the obtained policies on the evaluation data \( \mathcal{S}^{\text{FNN}}_{\text{evaluation}} \).

\subsection{Policy Learning Results under Continuous Policy Class}\label{sec:real_data_synthetic}
In this section, we present the results of policy learning using semi‐synthetic data. To evaluate the robustness of each method’s performance, we hold the training set 
$\mathcal{S}^{\mathrm{FNN}}_{\mathrm{train}}$
fixed and draw 30 bootstrap resamples (with replacement) from the inference set 
$\mathcal{S}^{\mathrm{FNN}}_{\mathrm{inference}}$ and evaluation set $\mathcal{S}^{\mathrm{FNN}}_{\mathrm{evaluation}}$.
As in Table~\ref{tab:realdata_regret_result}, Panel A of Table~\ref{tab:semi_syndata_regret_result} displays the mean of MPR over the 30 resamples for each benchmark method across policy classes.  The primary distinction is that, in this setting, the policy classes include
not only discrete treatment levels but also continuous policy classes. In the rightmost column of Table~\ref{tab:semi_syndata_regret_result}, we report the largest $p$-value obtained from pairwise comparisons of \textsf{DLPT} regret against each benchmark. Consistent with our findings on real data under discrete policies, the proposed \textsf{DLPT} framework significantly outperforms all benchmarks across all four policy classes (with the largest $p$-value $<0.05$), underscoring its superiority even in settings with continuous treatment regimes.

\begin{table}[!t]
\centering \scriptsize
\caption{Performance Comparison between Different Benchmarks}
\setlength{\tabcolsep}{0.4mm}{
\begin{tabular}{ccccccccccccccc}
\\[-1.8ex]\hline 
\hline \\[-1.8ex]
\multicolumn{15}{c}{Panel A: MPR Comparison of CATE of Different Benchmarks} \\
\hline \\[-1.8ex]
\multicolumn{2}{c}{Policy Class}&\multicolumn{2}{c}{\sf LR} & \multicolumn{2}{c}{\sf LogR} & \multicolumn{2}{c}{\sf LCM} & \multicolumn{2}{c}{\sf PDL} &\multicolumn{2}{c}{\sf UNI}   &\multicolumn{2}{c}{\sf DLPT}&Max $p$-value\\ 
&&\multicolumn{2}{c}{(1)}&\multicolumn{2}{c}{(2)}&\multicolumn{2}{c}{(3)}&\multicolumn{2}{c}{(4)}&\multicolumn{2}{c}{(5)}&\multicolumn{2}{c}{(6)}&\\[-1.8ex] \\
\hline  \\[-1.8ex] 
\multicolumn{2}{c}{Finite Observed} & \multicolumn{2}{c}{29.89\%}&\multicolumn{2}{c}{33.18\%}  & \multicolumn{2}{c}{31.04\%} &\multicolumn{2}{c}{27.58\%} &\multicolumn{2}{c}{28.00\%}&\multicolumn{2}{c}{17.96\%} &$4.65e^{-16}$ \\
\multicolumn{2}{c}{Discretized Continuous}&\multicolumn{2}{c}{33.04\%}&\multicolumn{2}{c}{34.75\%}&\multicolumn{2}{c}{33.44\%}&\multicolumn{2}{c}{29.28\%}&\multicolumn{2}{c}{32.80\%}&\multicolumn{2}{c}{19.69\%}&$2.11e^{-10}$\\
\multicolumn{2}{c}{Linear}&\multicolumn{2}{c}{6.01\%}&\multicolumn{2}{c}{6.37\%}&\multicolumn{2}{c}{6.44\%}&\multicolumn{2}{c}{0.66\%}&\multicolumn{2}{c}{1.75\%}&\multicolumn{2}{c}{0.01\%}&$3.86e^{-04}$\\
\multicolumn{2}{c}{DNN}&\multicolumn{2}{c}{12.65\%}&\multicolumn{2}{c}{9.20\%}&\multicolumn{2}{c}{13.33\%}&\multicolumn{2}{c}{7.42\%}&\multicolumn{2}{c}{8.46\%}&\multicolumn{2}{c}{0.99\%}&$1.95e^{-06}$\\
\hline \hline\\[-1.8ex]
\multicolumn{15}{c}{Panel B: APE Comparison of Learned Policy under DNN Policy Class within User Groups}\\
 \hline \\[-1.8ex] 
Group & $t^*$&\multicolumn{2}{c}{\sf LR} & \multicolumn{2}{c}{\sf LogR} & \multicolumn{2}{c}{\sf LCM} & \multicolumn{2}{c}{\sf PDL} &\multicolumn{2}{c}{\sf UNI}   &\multicolumn{2}{c}{\sf DLPT}&Max $p$-value\\ 
 && mean & APE& mean & APE& mean & APE& mean & APE& mean & APE& mean & APE&\\
 & & (1a)        & (1b)      & (2a)          & (2b)        & (3a)         & (3b)         & (4a)         & (4b)         & (5a)         & (5b)    & (6a)         & (6b)     &   \\[-1.8ex] \\
\hline  \\[-1.8ex] 
LL (35.92\%)& 443.31 &878.01&98.06\%&707.45&59.58\%&760.74&71.56\%&555.23&25.19\%&178.00&59.84\%&359.24&\textbf{18.96\%}&$4.79e^{-47}$\\
LH (32.08\%)& 354.76 &556.55&56.88\%&444.30&25.25\%&623.21&75.99\%&387.22&9.15\%&178.00&49.83\%&320.82&\textbf{9.56\%}&$1.69e^{-32}$\\
HL (24.76\%)&284.24 &188.28&33.76\%&196.51&39.87\%&422.51&48.77\%&110.46&61.14\%&178.00&37.38\%&247.66&\textbf{13.02\%}&$6.38e^{-57}$\\
HH (23.38\%)& 230.28 &77.34&66.42\%&98.96&57.09\%&311.06&35.22\%&63.42&72.46\%& 178.00&22.70\%&233.83&\textbf{1.51\%}&$1.26e^{-61}$\\
\hline\hline \\[-1.8ex]
\multicolumn{15}{c}{Panel C: Regret Comparison of Learned Policy Value under DNN Policy Class within User Groups}\\
 \hline \\[-1.8ex] 
Group & $V^*$&\multicolumn{2}{c}{\sf LR} & \multicolumn{2}{c}{\sf LogR} & \multicolumn{2}{c}{\sf LCM} & \multicolumn{2}{c}{\sf PDL} &\multicolumn{2}{c}{\sf UNI}   &\multicolumn{2}{c}{\sf DLPT}&Max $p$-value\\ 
 && mean & PR& mean & PR& mean & PR& mean & PR& mean & PR& mean & PR&\\
 & & (1a)        & (1b)      & (2a)          & (2b)        & (3a)         & (3b)         & (4a)         & (4b)         & (5a)         & (5b)    & (6a)         & (6b)     &   \\[-1.8ex] \\
\hline  \\[-1.8ex] 
LL (35.92\%) & 0.0432 &0.0339&21.52\%&0.0386&10.65\%&0.0330&23.61\%&0.0393&9.03\%&0.0383&11.34\%&0.0420&\textbf{2.78\%}&$1.21e^{-04}$\\
LH (32.08\%)& 0.0759 &0.0652&14.10\%&0.0694&8.56\%&0.0625&17.65\%&0.0699&7.91\%&0.0694&8.56\%&0.0758&\textbf{0.01\%}&$6.01e^{-07}$\\
HL (24.76\%)& 0.1645 &0.1471&10.58\%&0.1513&8.02\%&0.1436&12.71\%&0.1574&4.32\%&0.1564&4.92\%&0.1642&\textbf{0.02\%}&$3.81e^{-11}$\\
HH (23.38\%)& 0.1742 &0.1654&5.05\%&0.1672&4.02\%&0.1614&7.35\%&0.1697&2.58\%&0.1678&3.67\%&0.1741&\textbf{0.01\%}&$1.93e^{-14}$\\
\hline \hline\\[-1.8ex]
    \end{tabular}}
    \label{tab:semi_syndata_regret_result}
    \begin{tablenotes}
    \item Note: In Panels B and C, the percentage of each user group is reported in parentheses. Column $t^*$ in Panel B denotes the average optimal policy under the DNN policy class for each user group. Column $V^*$ in Panel C denotes the average optimal policy value under the DNN policy class within the user group. To preserve data confidentiality, this value is rescaled by multiplying it by a constant factor.
\end{tablenotes}
\end{table}


Now, we would like to provide some intermediate  steps for the policy learning  to investigate why our method could outperform other methods. Specifically, similar to Section~\ref{sec:HTE}, we divide users into four segments based on their production and consumption values prior to the experiment: low production combined with low consumption (LL), low production combined with high consumption (LH), high production combined with low consumption (HL), and high production combined with high consumption (HH). In Panels B and C of Table~\ref{tab:semi_syndata_regret_result}, we report the results of policy learning under the DNN policy class across methods.
For each group, we compute the ground-truth optimal point level $t^*$ under the DNN policy class and the corresponding ground-truth optimal value $V^*$. 
 
As shown in Panels B and C of Table~\ref{tab:semi_syndata_regret_result}, the optimal treatment levels and the corresponding optimal values vary considerably across user groups. For users in the LL group, the optimal ground-truth treatment level is $443.31$ with $0.0432$ as the average value, while for users in the HH group, it is only $230.28$ with $0.1742$ as the average value. Moreover, the optimal treatment level increases from HH user to LL user monotonically, while the optimal value decreases from HH user to LL user monotonically. This aligns with intuition, as more engaged users typically require fewer incentives to produce content, resulting in a lower optimal treatment level and higher optimal value. All methods capture this monotonic trend to some extent, but most fail to accurately estimate the true optimal levels, especially for the HL and HH groups. These two groups, although requiring lower incentives, still need meaningful treatment, yet many benchmark methods recommend treatment levels that are too low. For instance, {\sf UNI} always selects the minimum level $178$ regardless of the user group, resulting in a completely non-personalized policy. 
In contrast, our proposed {\sf DLPT} framework not only recovers the correct trend but also closely matches the true optimal policies across all groups, achieving significantly lower APEs and lower PRs (with the largest $p$-value$<0.05$). While {\sf PDL} performs reasonably well in distinguishing between groups, it struggles with HL and HH users, likely due to its limited ability to handle policies near treatment boundaries. We believe that the debiasing procedure used in {\sf DLPT} provides robustness when estimating optimal policies close to boundary levels (in our case, at levels $178$ or $738$), enabling more accurate personalization for these challenging cases.

\subsection{Regret Visualization over Sample Sizes}\label{sec:robustness}
\begin{figure}[!t]
    \centering
    \includegraphics[width=1\linewidth]{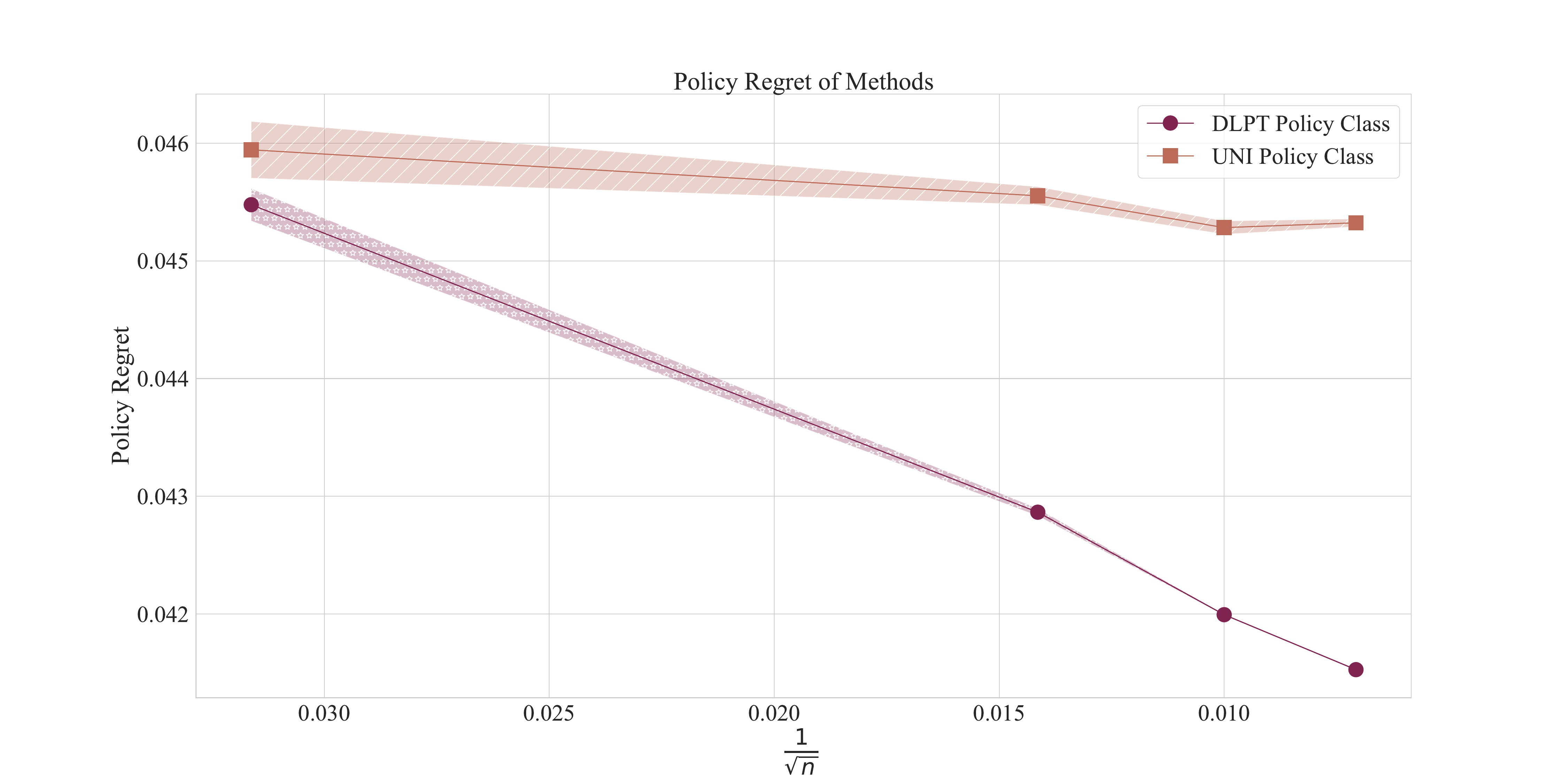}
    \caption{Policy Regret Scales with $1/\sqrt{n}$ }
    \label{fig:convergence}
\end{figure}
Next, we validate the regret bound derived from Theorem \ref{thm:learning_1} in Section~\ref{sec:theo_framework} using our data. Specifically, we first draw and fix a single validation set of $20,000$ observations from the user population for evaluation across all experiments. For each sample size $n \in $ \{$1,000$, $5,000$, $10,000$, $20,000$\}, 
we independently draw \( n \) observations to estimate nuisance parameters and another \( n \) observations to train the policy. We repeat this procedure 30 times for each \( n \), computing policy regret on the fixed validation set for each replication under DNN policy class. Figure \ref{fig:convergence} presents the policy regret of {\sf UNI}, {\sf PDL}, and {\sf DLPT} methods with the increase of $n$.

We highlight that as the training dataset size increases, the average policy regret of our {\sf DLPT} decreases linearly with respect to the square root of the sample size. This empirical pattern aligns precisely with our theoretical result in Theorem~\ref{thm:learning_1}, which predicts that policy regret scales as $1/\sqrt{n}$. Thus, our theoretical guarantees are not only rigorous but also empirically relevant in practical settings. This finding underscores that experimenters can reliably employ our method across experiments of varying sample sizes.

\subsection{Robustness Checks}\label{sec}
\begin{figure}[!t]
    \centering
    \includegraphics[width=1\linewidth]{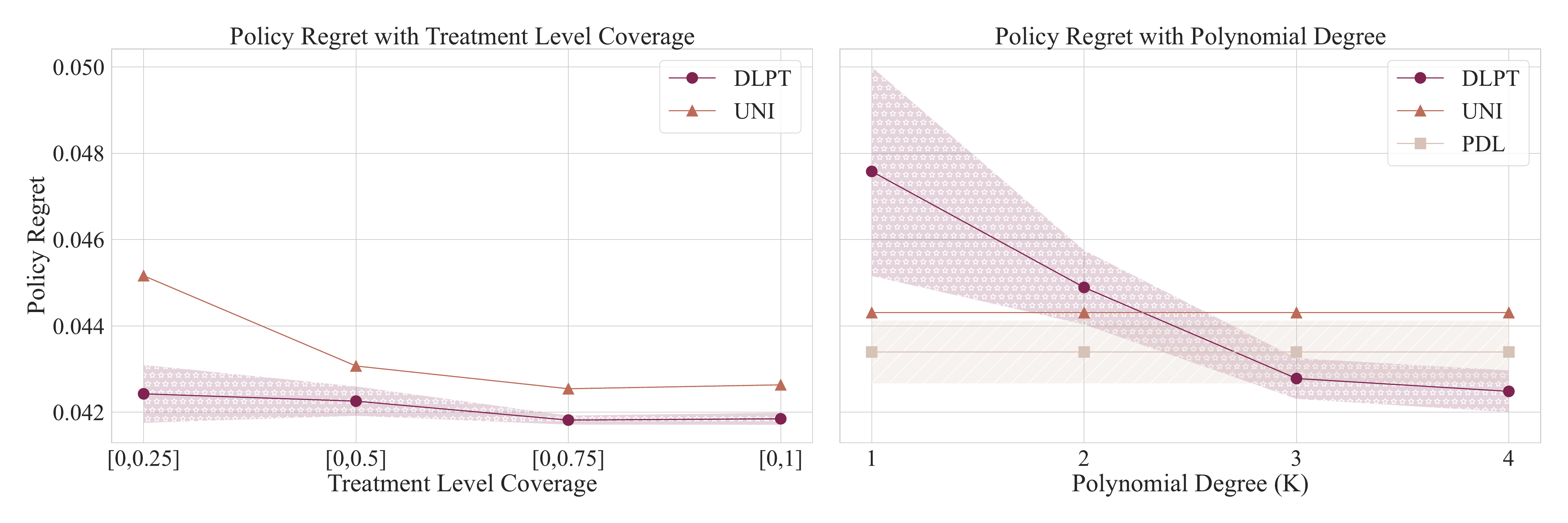}
    \caption{Policy Regret with Treatment Level Coverage and Polynomial Degree Assumption.}
    \label{fig:robustness}
\end{figure}

We also perform robustness checks to assess the resilience of the {\sf DLPT} framework under two practical challenges. Owing to space constraints, full details are deferred to Appendix~\ref{app:synthetic}, and here we provide a high‐level summary. First, we vary treatment coverage by restricting the span of treatment levels within the bounded treatment space. These experiments demonstrate that wider treatment coverage consistently enhances policy learning performance.
Second, we evaluate the performance under different assumed polynomial degrees \( K \in \{1, 2, 3, 4\} \). The resulting regret values align with the theoretical insights discussed in Section~\ref{sec:approximation_power}, confirming that the regret scales with \( K^{-s_t} \). Figure \ref{fig:robustness} presents the policy regret with treatment level coverage and polynomial degree assumption.


\section{Conclusion}\label{sec:conclusion}
In this paper, we introduce the {\sf DLPT} framework, a novel approach for estimating continuous policy values and learning personalized continuous policies with observations collected at discrete treatment levels. We provide theoretical guarantees for {\sf DLPT}, including the asymptotic unbiasedness of the policy value estimator and  $\sqrt{n}$-regret bound for the learned policies. The practical efficacy of {\sf DLPT} was rigorously tested through a large-scale field experiment on a leading video-sharing platform, where it demonstrated substantial improvements in policy value estimation and policy regret minimization over conventional methods. 

We conclude by outlining two avenues for future research. First, 
our analysis implicitly assumes that the Stage-3 empirical risk-minimization problem can be solved exactly, but we do not provide theoretical conditions that guarantee its computational tractability. 
Characterizing when---and under what assumptions---this optimization problem is solvable remains a significant open question, with broader implications for the policy learning literature beyond the scope of this work.
Second, while our framework focuses on identifying the optimal policy given a predefined set of discrete treatments,
it  also has the potential to inform  optimal experimental design --- specifically, the selection of  treatment levels to include in the experiment. Addressing this challenge would allow  practitioners to better integrate experimental design with downstream policy learning.  Although we offer   heuristic guidance in this paper,  such as increasing the parameter span tested, a more systematic investigation remains a promising future direction. 


\bibliographystyle{informs2014}
\bibliography{main}

\newpage
\renewcommand{\theHsection}{A\arabic{section}}
\begin{APPENDICES}
\begin{center}
\textbf{\Large Online Appendices}
\end{center}

\section{Technical Details}
\subsection{Assumptions}\label{app:assumptions}
\begin{assumption}\label{ass:loss function}
Following Assumption 1 in \cite{farrell2020deep}, we require Lipschitz continuity and sufficient curvature, near the truth, for loss function $\ell$ in Stage 1 nuisance parameter estimation.
    \begin{itemize}
        \item[(a)] {\sc (Lipschitz Continuity)} There exists a positive constant $C_{\ell}$ such that, for any $\bm \theta(\cdot)$, $\tilde{\bm \theta}(\cdot)$ and $\bmx$,
    \begin{equation}\label{eq:Lipschitz}
    \,|\,\ell(y,t,\bmtheta(\bmx))-\ell(y,t,\tilde{\bm{\theta}}(\bmx))|\leq C_{\ell}\|\bm{\theta}(\bmx)-\tilde{\bm{\theta}}(\bmx)\|_2,    
    \end{equation}
     \item[(b)] {\sc (Sufficient Curvature)} There exist positive constants $c_1$ and $c_2$ such that, for any $\bmtheta(\cdot)\in\mathcal{F}_{DNN}$,
    \begin{equation}\label{eq:curvature}
    c_1\E[\|\bm{\theta}(\bmX)-{\bm{\theta}}^*(\bmX)\|_2^2]\leq \E[\ell(Y,T,\bm{\theta}(\bm{X}))]- \E[\ell(Y,T,\bm{\theta}_K^*(\bm{x}))]\leq c_2\E[\|\bm{\theta}(\bmX)-{\bm{\theta}}^*(\bmX)\|_2^2].    
    \end{equation}
    \end{itemize}
\end{assumption}

\begin{assumption}\label{ass:data bounded and nuisance smoonth}
    Following Assumption 2 in \cite{farrell2020deep}, we make the regularity assumptions on samples and nuisance parameter function $\bm\theta^*(\bm x)$:
    \begin{itemize}
    \item[\textit{(a)}] $\bm{Z}_i=(\bm{X}_i',T_i,Y_i)'$, $1\leq i\leq n$, are i.i.d. copies from the population random variables $\bm{Z}=(\bm{X}',T,Y)'\in [-1,1]^{d_{\bmX}}\times \mathcal{T}\times \mathcal{Y}$, where $\mathcal{Y}$ is the bounded support of the outcome $Y$.
    \item[\textit{(b)}] The parameter function $\bm{\theta}_K^*(\bm{x})$ is uniformly bounded. Furthermore, $\theta^*_{k}(\bm{x})\in W^{p,\infty}([-1,1]^{d_{\bmX}})$, $k=1,2,\dots,d_{\bmtheta}$, where for positive integers $p$, define the 
    Sobolev ball $W^{p,\infty}([-1,1]^{d_{\bmX}})$ of functions $h:\R^{d_{\bmX}}\mapsto\R$ with smoothness $p\in\mathbb{N}_+$ as,
    $$W^{p,\infty}([-1,1]^{d_{\bmX}}):=\Big\{h:\max_{\bm{r},|\bm{r}|<p} \esssup_{\bm{v}\in[-1,1]^{d_{\bmX}}} |D^{\bm{r}} h(\bm{v})|\leq 1 \Big\},$$
    where $\bm{r}=(r_1,\dots,r_{d_{\bmX}}),~|\bm{r}|=r_1+\dots+r_{d_{\bmX}}$ and $D^{\bm{r}} h$ is the weak derivative.
    \end{itemize}
\end{assumption}

\begin{assumption}\label{ass:influence function}
    The following regularity conditions hold uniformly on the distribution of $\bm Z=(\bm X',T,Y)'$:
    \begin{itemize}
        \item[(a)] $\bm \Lambda(\bm x)=\bbE[\ell_{\bm\theta\bm\theta}(y,t,\bm\theta(\bm x))|\bm X = \bm x]$ is invertible with bounded inverse. Specifically, $\bm \Lambda(\bm x)=\bbE[G_{\bm \theta}(\hat{\bm\theta}(\bm x),t)G_{\bm \theta}((\hat{\bm\theta}(\bm x),t)'|\bm X= \bm x]$ if using MSE as loss function in Stage 1 and $\bm \Lambda(\bm x)=\bbE[G_{\bm \theta}(\hat{\bm\theta}(\bm x),t)\tilde{\bm T}_K'|\bm X= \bm x]$ if using binary cross entropy as loss function in Stage 1.
        \item[(b)] Value function $H(\bm\theta(\bm x),\pi(\bm x))$ is identified and pathwise differentiable.
    \end{itemize}
\end{assumption}

\begin{assumption}
\label{ass:true_dgp_smooth}
 {\sc (H\"older Continuity of Latent Function)} 
Consider the oracle DGP $f:\mathcal{X}\times\mathcal{T}\rightarrow[0,1]$ such that $f(\bm x,t)=\mathbb{E}[Y|\bm X=\bm x,T=t]= \sigma(\tilde{f}(\bm{x}, t))$,
where $\sigma(z) = \frac{1}{1 + e^{-z}}$ is the sigmoid function, and $\tilde{f} : \mathcal{X} \times \mathcal{T} \to \mathbb{R}$ is a latent function. It satisfies that, for all $\bm x\in\mathcal{X}$ , the latent function $t\rightarrow \tilde{f}(\bm x ,t)$ is $s_t$-H\"older continuous with a constant  $L>0$  that does not depend on $\bm x$. Specifically, 
     there exist $L>0$ and $s_t$ such that for all $\bm x\in\mathcal{X}$ and for all $t,t'\in \mathcal{T}$, 
    \begin{equation}
        |\tilde{f}(\bm x,t)-\tilde{f}(\bm x,t')|\le L|t-t'|^{s_t},
    \end{equation}
    where $s_t$ (with $0\le s_t\le 1$ typically) measures the smoothness of $\tilde{f}(x,\cdot)$.
\end{assumption}

\subsection{Proof of Proposition \ref{prop: nuisance convergence_rate}}\label{appendix:prop convergence rate proof}

 We first prove the nonparametric identifiability of nuisance parameter $\bm\theta^*(\bm x)$ in Proposition \ref{prop: nuisance convergence_rate}(a). Following DGP defined in Equation \ref{eq:DGP}, let $G(\bm\theta^*(\bm x),t)=G(\tilde{\bm\theta}(\bmx),t)$. To show $\bm \theta^*(\bm x)$ can be nonparametrically identified, we show that $\bm\theta^*(\bm x)=\tilde{\bm \theta}(\bm x)$.
  Since  $G(\bm\theta^*(\bm x),t)=G(\tilde{\bm\theta}(\bmx),t)$, and the sigmoid function is reversible, we have that
  \begin{gather*}
      \theta^*_0(\bm x)+\theta^*_1(\bm x)t+\theta^*_2(\bm x)t^2+\theta^*_3(\bm x)t^3=\tilde{\theta}_0(\bm x)+\tilde{\theta}_1(\bm x)t+\tilde{\theta}_2(\bm x)t^2+\tilde{\theta}_3(\bm x)t^3\\
     ( \bm\theta^*(\bm x)-\tilde{\bm\theta}(\bm x))(1,t,t^2,t^3)'=0
  \end{gather*}
  Define $\tilde{\bm T}= (1,t,t^2,t^3)$, we therefore have
  \begin{gather*}
      ( \bm\theta^*(\bm x)-\tilde{\bm\theta}(\bm x))\tilde{\bm T}'=0\\
      \bbE[(\bm\theta^*(\bm X)-\tilde{\bm\theta}(\bm X)\tilde{\bm T})^2|\bm X] = (\bm\theta^*(\bm X)-\tilde{\bm\theta}(\bm X))\bbE[\tilde{\bm T}\tilde{\bm T}'|\bm X] (\bm\theta^*(\bm X)-\tilde{\bm\theta}(\bm X))'=0
  \end{gather*}
  Since $\bbE[\tilde{\bm T}\tilde{\bm T}'|\bm X] $ is positive definite uniformally with respect to $\bm X$, we have that $\bm\theta^*(\bm X)-\tilde{\bm\theta}(\bm X)=0$, i.e., $\bm\theta^*(\bm x)$ can be nonparametrically identified.

  We proceed to prove the convergence rate of nuisance parameter estimators in Proposition \ref{prop: nuisance convergence_rate}(b). We use the following result in \cite{farrell2020deep}.
\begin{lemma}[Theorem 1 in \cite{farrell2020deep}]\label{lemma:DNN_converge}
Suppose Assumptions \ref{ass:loss function} and \ref{ass:data bounded and nuisance smoonth} hold.
With the structured DNN of width $H=O(n^{\frac{d_{\bmX}}{2(p+d_{\bmX})}}\log^2 n)$ and depth $L=O(\log n)$, there exists a constant $C$ such that 
\begin{equation}\nonumber
\|\hat{\bmtheta}_k-\bmtheta^*_k\|_{L_2(\bm{X})}^2~\lesssim~ 
n^{-\frac{p}{p+d_{\bmX}}}\log^8 n+\frac{\log\log n}{n}
\end{equation}
and 
\begin{equation}\nonumber
\E_n\big[(\hat{\bmtheta}_k-\bmtheta^*_k)^2\big]~\lesssim~ 
n^{-\frac{p}{p+d_{\bmX}}}\log^8 n+\frac{\log\log n}{n}
\end{equation}
for $n$ large enough with probability at least $1-\exp(n^{-\frac{d_{\bmX}}{p+d_{\bmX}}}\log^8 n)$, for $k=1,\dots,d_{\bmtheta}$.
\end{lemma}
To achieve the convergence rate in Proposition \ref{prop: nuisance convergence_rate}(b), we need to verify assumptions made in Lemma \ref{lemma:DNN_converge}. 
\begin{enumerate}
    \item The nonparametric identifiability of $\bm\theta^*(\bm x)$ can be satisfied as long as  $\bbE[\tilde{\bm T}\tilde{\bm T}'|\bm X] $ is positive definite uniformly with respect to $\bm X$, which is satisfied in our RCT setting when the number of discrete levels of treatment $m>=4$. 
    \item The Lipschitz continuity of loss function $\ell$ is satisfied under either MSE loss function or binary cross-entropy loss function since $G(\cdot)$ is a sigmoid function with bounded random variable $\bm Z$.
    \item The curvature condition of the loss function near the truth can be verified through the following derivations.
    When using MSE as loss function:
\begin{gather*}
    \bbE[\ell(Y,T,\bm{\theta}(\bm{X}))]- \bbE[\ell(Y,T,\bm{\theta}_K^*(\bm{X}))] =
  \bbE\left[
   \left(G(\bmtheta(\bmX),T) - G(\bmtheta^*(\bmX),T)\right)^2
  \right]
\end{gather*}
Since $G(\cdot)$ is a continuous and differentiable with respect to $\bm \theta(\bm x )$, we can find $\bm \theta_0(\bm x)$ which has $\theta_{0k}(\bm x)\in(\theta_{k}(\bm x),\theta^*_{k}(\bm x))$ for each $k$ satisfies that $G_{\bm\theta}(\bm \theta_0(\bm x),t)=\frac{G(\bm \theta(\bm x),t) - G(\bm \theta^*(\bm x),t)}{\bm \theta(\bm x)-\bm\theta^*(\bm x)}$. Therefore, we have
\begin{align}
         \bbE[\ell(Y,T,\bm{\theta}(\bm{X}))]- \bbE[\ell(Y,T,\bm{\theta}_K^*(\bm{x}))] &=
     \bbE\left[(\bm\theta(\bm X)-\bm\theta^*(\bm X))G_{\bm\theta}(\bm\theta_0(X),T)G_{\bm\theta}(\bm\theta_0(X),T) ' (\bm\theta(\bm X)-\bm\theta^*(\bm X))'\right]\label{eq:second ineq}\\
     &=\bbE[((\bm\theta(\bm X)-\bm\theta^*(\bm X))\tilde{\bm T})^2G(\bm\theta_0(X),T)^2(1-G(\bm\theta_0(X),T))^2]\label{eq:first ineq}
\end{align}
From Equation \eqref{eq:second ineq}, we verify that the second inequality in the curvature condition is satisfied since 
$G(\cdot)$ is bounded away from 0 and 1 by the boundedness of the covariate space and treatment space. Deriving from Equation \eqref{eq:first ineq}, we further have that
\begin{align*}
     \bbE[\ell(Y,T,\bm{\theta}(\bm{X}))]- \bbE[\ell(Y,T,\bm{\theta}_K^*(\bm{x}))] &\ge c\bbE[((\bm\theta(\bm X)-\bm\theta^*(\bm X))\tilde{\bm T})^2]\\
     &\ge d \bbE[ (\bm\theta^*(\bm X)-\tilde{\bm\theta}(\bm X))\bbE[\tilde{\bm T}\tilde{\bm T}'|\bm X] (\bm\theta^*(\bm X)-\tilde{\bm\theta}(\bm X))']\\
     &\ge e\bbE[ (\bm\theta^*(\bm X)-\tilde{\bm\theta}(\bm X)(\bm\theta^*(\bm X)-\tilde{\bm\theta}(\bm X))']
\end{align*}
which proves that the first inequality in the curvature condition is satisfied.
\end{enumerate}
To this end, we conclude the proof for Proposition \ref{prop: nuisance convergence_rate}.
\subsection{Proof of Proposition \ref{prop:influence function}}\label{appendix:infl func proof}
We derive the influence function in Proposition  \ref{prop:influence function} following Theorem 2 in \cite{farrell2020deep}.
\begin{lemma}[Theorem 2 in \cite{farrell2020deep}]\label{lemma:influncefunc}
    The following conditions hold uniformly in the given conditioning elements. \textit{(i)} DGP \eqref{eq:DGP} holds and identifies $\bmtheta^*(\cdot)$. \textit{(ii)} $\E[\ell_{\bmtheta}(y,t,\bmtheta^*(\bmx))|\bmX=\bmx,T=t]=0$. \textit{(iii)} $\bm{\Lambda}(x):=\E[\ell_{\bmtheta\bmtheta}(Y,T,\bmtheta(\bmx))|\bmX=\bmx]$ is invertible with uniformly bounded inverse. \textit{(iv)} Parameter $V_K(\pi)$ is identified, pathwise differentiable, and $H$ and $\ell$ are thrice continuously differentiable in $\bmtheta$. \textit{(v)} $H(\bmtheta^*(\bmX),\pi(\bmx))$ and $\ell_{\bmtheta}(Y,T,\bmtheta^*(\bmX))$ possess $q>4$ finite absolute moments and positive variance. Then for the policy value $Q(\pi)$, the Neyman orthogonal score is $\psi(\bm{z},\pi;\bm \theta,\bm \Lambda)-V_K(\pi)$, where
\begin{equation}
    \psi(\bm z,\pi;\hat{\bm\theta},\bm\Lambda) = H(\hat{\bm\theta}(\bm x),\pi(\bm x)) - H_{\bm\theta}(\hat{\bm\theta}(\bm x), \pi(\bm x))\bm\Lambda(\bm x)^{-1} \ell_{\bm\theta}(y,  t, \hat{\bm\theta}(\bm x)),
\end{equation}
where $\ell_{\bmtheta},H_{\bmtheta}$ are $d_{\bmtheta}$-dimensional vectors of first order derivatives, and $\ell_{\bmtheta\bmtheta}$ is the $d_{\bmtheta}\times d_{\bmtheta}$ Hessian matrix of $\ell$, with $\{k_1,k_2\}$ element defined by $\partial^2\ell/\partial \theta_{k_1}\partial \theta_{k_2}$.
\end{lemma}
To obtain the influence function for {\sf DLPT} policy value estimator $\hat{V}^{\sf DLPT}_K(\pi)$, we need to verify assumptions made in Lemma \ref{lemma:influncefunc}.
\begin{enumerate}
    \item The nonparametric identifiability of $\bm\theta^*(\bm x)$ can be satisfied as long as  $\bbE[\tilde{\bm T}\tilde{\bm T}'|\bm X] $ is positive definite uniformly with respect to $\bm X$, which is satisfied in our RCT setting when the number of discrete levels of treatment $m>=4$.
    
    \item  $\ell_{\bm\theta}(y,t,\bmtheta^*(\bmx))=G_{\bm \theta}(\bmtheta^*(\bmx),t)(G(\bmtheta^*(\bmx),t)-y)$ if using MSE as loss function in Stage 1, and $\ell_{\bm\theta}(y,t,\bmtheta^*(\bmx))=(G(\bmtheta^*(\bmx),t)-y)\tilde{T}$ if using binary cross entropy as loss function in Stage 1. (ii) holds under both loss functions with DGP assumption in Equation \eqref{eq:DGP}.
    
    \item $\bm \Lambda(\bm x)=\bbE[G_{\bm \theta}((\hat{\bm\theta}(\bm x),t)G_{\bm \theta}((\hat{\bm\theta}(\bm x),t)'|\bm X= \bm x]$ if using MSE as loss function in Stage 1 and $\bm \Lambda(\bm x)=\bbE[G_{\bm \theta}((\hat{\bm\theta}(\bm x),t)\tilde{T}']$ if using binary cross entropy as loss function in Stage 1. (iii) holds under both loss functions as $G(\cdot)$ is bounded away from 0 and 1 with the boundedness of covariate and treatment spaces, as well as $G_{\bm \theta}=G(1-G)\tilde{T}$. 
    
    \item (iv) holds as both $H$ and $\ell$ enjoy sufficient smoothness.
    
    \item (v) holds as $G(\cdot)$ enjoys sufficient smoothness with bounded value of nuisance parameter.
\end{enumerate}
To this end, we conclude our proof of Proposition \ref{prop:influence function}.
\subsection{Proof of Proposition \ref{prop:orthogonality}}\label{appendix:orthogonal proof}
To show the universal orthogonality, we conduct the following derivation:
    \begin{align*}
          &\bbE[\nabla_{\bm \theta}\psi(\bm z,\pi;\bm \theta^*, \bm \Lambda)|\bm X = \bm x]\\
          =& \bbE\Big[H_{\bm\theta}(\bm\theta(\bm x),\pi(\bm x)) -H_{\bm\theta}(\bm\theta(\bm x),\pi(\bm x))\bm \Lambda(\bm x)^{-1} \ell_{\bm\theta\bm\theta}(y,t,\bm \theta(\bmx)) - H_{\bm\theta\bm\theta}(\bm\theta(\bm x),\pi(\bm x))\bm \Lambda(\bm x)^{-1} \ell_{\bm\theta}(y,t, \bm\theta(\bmx)) \Big| \bm X=\bm x\Big]\\
           \stackrel{(i)}{=} &\bbE\Big[- H_{\bm\theta\bm\theta}(\bm\theta(\bm x),\pi(\bm x))\bm \Lambda(\bm x)^{-1} \ell_{\bm\theta}(y,t, \bm\theta(\bmx)) \Big| \bm X=\bm x\Big]\\
            = &-  H_{\bm\theta\bm\theta}(\bm\theta(\bm x),\pi(\bm x))\bm \Lambda(\bm x)^{-1}\bbE\Big[ \ell_{\bm\theta}(y,t, \bm\theta(\bmx)) \Big| \bm X=\bm x\Big]\stackrel{(ii)}{=}0 
    \end{align*}
    where (i) is by definition that $\bm{\Lambda}(x):=\E[\ell_{\bmtheta\bmtheta}(Y,T,\bmtheta(\bmx))|\bmX=\bmx]$ , and (ii) is by the first order optimality of $\bm\theta(\cdot)$.

\subsection{Proof of Theorem \ref{thm:learning_1}}\label{appendix:regret bound proof}
We first prove a more involved version of the regret bound, which will imply Theorem \ref{thm:learning_1} as shown in Appendix~\ref{appendix:prove_learning_1}. To establish this new regret bound, below we introduce a few more technical terms to characterize the policy class complexity:
\begin{definition}[Covering Number and Entropy Integral]
   For a function $f:\calX\rightarrow R$, denote its $L_2$-norm as $\|f\|_2=\sqrt{\bbE[f(x)^2]}$ and its empirical $L_2$-norm as $\|f\|_{2,n}=\sqrt{\bbE_n[f(x)^2]}$.  For a real-valued function class $\calF$, define the following properties to measure its complexity:
    \begin{itemize}
        \item Empirical covering number $\calN_2(\epsilon, \calF, x_{1:n})$: the  size of the smallest function class $\calF_\epsilon$, where $\calF_\epsilon$ is the $\epsilon$-cover of $\calF$ such that for any $f\in\calF$, there exists an $f_\epsilon\in\calF_\epsilon$ that satisfies $\|f-f_\epsilon\|_{2,n}\leq \epsilon$.
        \item  Covering number $\calN_2(\epsilon,\calF, n)$:  the maximal empirical covering number over all possible $n$ samples. 
        \item Entropy integral $  \kappa(r, \calF) = \inf_{\alpha\geq 0}\bigg\{
            4 \alpha + 10\int^r_{\alpha}\sqrt{\frac{\log\mathcal{N}_2(\epsilon,\mathcal{F},n)}{n}}d\epsilon 
            \bigg\}$.
    \end{itemize}
\end{definition}

We now  introduce the following regret bound.
\begin{proposition}
\label{prop:learning_2}
Let $S_1$ and $S_2$ be randomly and evenly splits of $n$ i.i.d.~samples. Let $\hat{\bm \theta}$ be the estimated nuisance on sample $S_1$ following procedure in Stage 1. Suppose that  $||\hat{\bm \theta}-\bm\theta^*||_{L_2(\bm X)} = o(n^{-1/4})$, which is guaranteed by Proposition \ref{prop: nuisance convergence_rate}. Let $\hat{\pi}$ be the learned policy that solves the  ERM problem: $ \hat{\pi}:=\inf_{\pi\in{\Pi}}~\{-\bbE_{S_2}[\psi(z,\pi;\hat{\bm \theta}, \Lambda)]\}$,
    where $\bbE_{S_2}[\cdot]$ denotes the empirical average over sample $S_2$. 
Define the function class: $\psi\circ \Pi = \{\psi(\cdot, \pi;\hat{\bm \theta}, {\bm \Lambda}):\pi\in\Pi\}.$
    Let $r=\sup_{\pi\in\Pi}\|\psi(\cdot, \pi;\hat{\bm \theta}, {\bm \Lambda}) - \psi(\cdot, \pi^*;\bm \hat{\bm \theta}, {\bm \Lambda}) \|_{L_2(\bm X)}$.    Assume Assumptions \ref{ass:loss function}, \ref{ass:data bounded and nuisance smoonth}, and \ref{ass:influence function} in Appendix \ref{app:assumptions} hold, then with probability $1-\delta$,
    \begin{equation}
        R_K(\hat{\pi}) = O\bigg(
         \kappa(r, \psi\circ \Pi) +   \frac{\log\calN_2(r,\psi\circ \Pi, n)}{n} 
        + r\sqrt{\frac{\log(1/\delta)}{n}} + \frac{\log(1/\delta)}{n}
        \bigg).
    \end{equation}
\end{proposition}

In the following, we shall first prove Proposition \ref{prop:learning_2} in Appendix \ref{appendix:prove_learning_2} and then prove Theorem \ref{thm:learning_1} in Appendix \ref{appendix:prove_learning_1}.
\subsubsection{Proving Proposition \ref{prop:learning_2}.}\label{appendix:prove_learning_2} The following two lemmas form a building block in our proof.

\begin{lemma}[\cite{foster2019orthogonal}, Theorem 2]\label{lemma:nuisance_learning} Let $S=\{(\bm z_i)_{i=1}^n\}$ be $n$ i.i.d.~observations sampled from space $\calZ$ under distribution $\calD$. Denote $\calF$ as the function class of the decision variable and $\calG$ as the  function class of a nuisance model. Let $\ell:\calZ\times\calF\times\calG\rightarrow R$ be a loss function.
Let $S_1$ and $S_2$ be two evenly and randomly splits of sample $S$. Let $\hat{g}\in\calG$ be the estimated nuisance on $S_1$ and $g^*\in\calG$ be the true nuisance parameter. Let $\hat{f} = \inf_{f\in\calF} \bbE_{S_2}[\ell(\bm z,f,\hat{g})]$ be the solution to the constrained ERM, where $\bbE_{S_2}[\cdot]$ denotes the sample average on sample $S_2$. 

Let $L_\calD(f,g)$ denote the population loss with respect to $(f,g)$ when marginalized over the sample space $\calZ$ under distribution $\calD$.
Suppose $L_\calD(f,g)$ satisfies universal orthogonality condition with respect to the nuisance parameter $g$. Suppose that  $L_\calD(f,g)$ are thrice-continuously differentiable and that $L_\calD(f,g)$ has bounded curvature with respect to $g$. 
Then the constrained ERM $\hat{f}$ has the following excess risk bound:
\begin{equation}
    L_\calD(\hat{f}, g^*) -     L_\calD(f^*, g^*) =O_p\big( \{L_\calD(\hat{f}, \hat{g}) -     L_\calD(f^*, \hat{g}) \} + \|\hat{g}-g^*\|_{L_2(\bm Z)}^2
    \big),
\end{equation}
where $f^*=\argmin_{f\in\calF}L_\calD(f, g^*)$.
\end{lemma}

Lemma \ref{lemma:nuisance_learning} shows that, with universal orthogonal loss functions, the regret bound can be decomposed into (i) the regret bound evaluated  at the estimated nuisance and (ii) the nuisance estimation error. The following result quantifies the first component.

\begin{lemma}[\cite{foster2019orthogonal}, Theorem 4; \cite{chernozhukov2019semi}, Theorem 4]\label{lemma:regret_bound_estimate}
Follow the notations in Lemma \ref{lemma:nuisance_learning}.
Define the function class  $\ell\circ \calF:=\{\ell(\theta(\cdot),\hat{g}_n(\cdot), \cdot):f\in\calF_n\}$.
Let $r=\sup_{f\in\calF}\|\ell(\cdot, f, \hat{g}) - \ell(\cdot, f^*, \hat{g}) \|_{L_2(\bm Z)}$. Then with probability $1-\delta$,
    \begin{equation}
     L_\calD(\hat{f}, \hat{g}) -     L_\calD(f^*, \hat{g}) = O\bigg(
         \kappa(r, \ell\circ \calF) +   \frac{\log\calN_2(r,\ell\circ \calF, n)}{n} 
        + r\sqrt{\frac{\log(1/\delta)}{n}} + \frac{\log(1/\delta)}{n}
        \bigg).
    \end{equation}
\end{lemma}

Recall that in our settings, $\psi()$ is universally orthogonal with respect to the nuisance $\bm \theta$, by Proposition \ref{prop:orthogonality}. We also assume that the nuisance component admits the following estimation rate: $||\hat{\bm \theta}-\bm\theta^*||_{L_2(\bm X)} = o(n^{-1/4})$, which is also characterized in Proposition \ref{prop: nuisance convergence_rate}. Moreover, Assumption \ref{ass:loss function} gives us the bounded curvature. Combining Lemma \ref{lemma:nuisance_learning} and Lemma \ref{lemma:regret_bound_estimate}, we have that with probability $1-\delta$,
\begin{align*}
    R_K(\hat{\pi}) & =O \bigg(
         \kappa(r, \psi\circ \Pi) +   \frac{\log\calN_2(r,\psi\circ \Pi, n)}{n} 
        + r\sqrt{\frac{\log(1/\delta)}{n}} + \frac{\log(1/\delta)}{n} + \|\hat{\bm \theta}-\bm\theta^*\|_{L_2(\bm X)}^2
        \bigg)\\
      &  =O\bigg(
         \kappa(r, \psi\circ \Pi) +   \frac{\log\calN_2(r,\psi\circ \Pi, n)}{n} 
        + r\sqrt{\frac{\log(1/\delta)}{n}} + \frac{\log(1/\delta)}{n}
        \bigg) + o(n^{-1/2})\\
    &  = O\bigg(
         \kappa(r, \psi\circ \Pi) +   \frac{\log\calN_2(r,\psi\circ \Pi, n)}{n} 
        + r\sqrt{\frac{\log(1/\delta)}{n}} + \frac{\log(1/\delta)}{n}
        \bigg),
\end{align*}
completing the proof.

\subsubsection{Proving Theorem \ref{thm:learning_1}.}\label{appendix:prove_learning_1}
Now we employ Proposition \ref{prop:learning_2} to show Theorem \ref{thm:learning_1}.
Consider $\psi\circ \Pi$ as a function class with VC subgraph dimension as $d$. Then we have
$\calN_2(\epsilon, \psi\circ \Pi, n) = O\big(\exp(d\left(1+\log(1/\epsilon)\right))\big) $ \citep{van2000asymptotic}. We therefore have
\begin{align*}
    \kappa(r, \psi\circ \Pi) &\leq 10\int_0^r\sqrt{\frac{d(1+\log(1/\epsilon))}{n}}\mbox{d}\epsilon\\
    &=10\epsilon\sqrt{\frac{d(1+\log(1/\epsilon))}{n}}\mid_0^r +10\int_0^r\sqrt{\frac{d}{n}}\frac{1}{2\sqrt{1+\log(1/\epsilon)}}\mbox{d}\epsilon\\
    &=r\sqrt{d/n}(1+\sqrt{1+\log(1/r)}).
\end{align*}
By Proposition \ref{prop:learning_2}, we have the regret bound as
    \begin{align*}
        R_K(\hat{\pi}) & = O\bigg(
         \kappa(r, \psi\circ \Pi) +   \frac{\calN_2(r,\psi\circ \Pi, n)}{n} 
        + r\sqrt{\frac{\log(1/\delta)}{n}} + \frac{\log(1/\delta)}{n}
        \bigg)\\
        & = O\bigg(
        r\sqrt{d/n}(1+\sqrt{1+\log(1/r)})+   \frac{d(1+\log(1/r)}{n} 
        + r\sqrt{\frac{\log(1/\delta)}{n}} + \frac{\log(1/\delta)}{n}
        \bigg)\\
        &= O\bigg(
        r(1+\sqrt{1+\log(1/r)})\sqrt{\frac{d}{n}}
        + r\sqrt{\frac{\log(1/\delta)}{n}} 
        \bigg)\\
        &=O\bigg(
        r(1+\sqrt{\log(1/r)})\sqrt{\frac{d}{n}}
        + r\sqrt{\frac{\log(1/\delta)}{n}} 
        \bigg),
     \end{align*}
completing our proof.

\subsection{Proof of Theorem \ref{prop:representation power}}\label{appendix:prop approximation proof}
We apply the classical Jackson's theorem \citep{jackson1912approximation} to provide the approximation error bound.
Since $\tilde{f}(\bmx,t)$ is $s_t$-H\"older continuous by Assumption \ref{ass:true_dgp_smooth} , by Jackson’s Theorem for univariate Hölder functions, we have that for each $\bm{x} \in \mathcal{X}$, there exists a univariate polynomial $P_K(t; \bm{x})$ of degree at most $K$ such that
$$
\sup_{t \in \mathcal{T}} \left| \tilde{f}(\bm{x}, t) - P_K(t; \bm{x}) \right| \leq C \cdot K^{-s_t},
$$
for some constant $C > 0$ depending on the H\"older constant and the domain $\mathcal{T}$. Define the approximating function
$G(\bm{\theta}^*(\bm{x}), t) := \sigma(P_K(t; \bm{x}))$,
where $\bm{\theta}_K^*(\bm{x}) \in \mathbb{R}^{p+1}$ are the coefficients of $P_K(\cdot; \bm{x})$.
Using the Lipschitz continuity of the sigmoid function $\sigma(\cdot)$ with constant $L_\sigma \leq \frac{1}{4}$, we obtain:
\[
\left| f(\bm{x}, t) - G(\bm{\theta}^*(\bm{x}), t) \right|
= \left| \sigma(\tilde{f}(\bm{x}, t)) - \sigma(P_K(t; \bm{x})) \right|
\leq L_\sigma \cdot \left| \tilde{f}(\bm{x}, t) - P_K(t; \bm{x}) \right|.
\]
Therefore,
\[
\sup_{(\bm{x}, t) \in \mathcal{X} \times \mathcal{T}} \left| f(\bm{x}, t) - G(\bm{\theta}^*(\bm{x}), t) \right| \leq \frac{C}{4} \cdot K^{-s_t} = O(K^{-s_t}),
\]
which completes the proof for the pointwise approximation error bound. For the policy value approximation error, we can rewrite it by the linearity of expectation,
\[
|V(\pi) - V_K(\pi)| \leq \mathbb{E}[w] \sup_{\bm{x} \in \mathcal{X}} \left| f(\bm{x}, \pi(\bm{x})) - G(\bm{\theta}^*(\bm{x}), \pi(\bm{x})) \right| = O(K^{-s_t}).
\]
Therefore, we obtain the bound
$|V(\pi) - V_K(\pi)| = O(K^{-s_t})
$
as required.

Last but not least, we 
Let $\pi^*_K = \arg\max_{\pi\in\Pi}V_K(\pi)$, and we have
\begin{align*}
    V(\pi^*)-V(\hat{\pi}_K) & = V(\pi^*)-V(\pi_K^*)+V(\pi_K^*)-V_K(\pi_K^*)+V_K(\pi_K^*)-V_K(\hat\pi_K)+V_K(\hat\pi_K)-V(\hat{\pi}_K)\\
    &\le V(\pi^*)-V(\pi_K^*)+|V(\pi_K^*)-V_K(\pi_K^*)|+|V_K(\pi_ K^*)-V_k(\hat\pi_K)|+|V_k(\hat\pi_K)-V(\hat{\pi}_K)|
\end{align*}
 We have that the second term and the last term in the right are uniformly bounded by $O(K^{-s_t})$. By Theorem \ref{thm:learning_1}, we have that the regret of the learned policy, third term on the right, is bounded by  $O\bigg(r\left(1+\sqrt{\log(1/r)}\right)\sqrt{\frac{d}{n}}+ r\sqrt{\frac{\log(1/\delta)}{n}}
        \bigg).$
For the first term of the right, 
We have:
\begin{align*}
    V(\pi^*) - V(\pi_K^*) = &\bc{V(\pi^*) - V_K(\pi^*)}  - V_K(\pi_K^*) + V(\pi_K^*) - \bc{ V(\pi_K^*)-V(\pi_K^*)}\\
    \stackrel{(i)}{\leq} & \bc{V(\pi^*) - V_K(\pi^*)}   - \bc{ V(\pi_K^*)-V(\pi_K^*)}\\
    \leq & 2 \sup_{\pi\in \Pi} |V(\pi)-V_K(\pi)| \stackrel{(ii)}{=} O\bigg(K^{-s_t}\bigg),
\end{align*}
where (i) is by the definition that $\pi_K^*$ optimizes $V_K(\cdot)$ and (ii) is by the uniform approximation error in Theorem \ref{prop:representation power}(b).
Therefore, we have 
\begin{gather*}
     V(\pi^*)-V(\hat{\pi}_K) = O\bigg(r\left(1+\sqrt{\log(1/r)}\right)\sqrt{\frac{d}{n}}+ r\sqrt{\frac{\log(1/\delta)}{n}} + K^{-s_t}
        \bigg)
\end{gather*}

\section{Empirical Validation}

\subsection{User Covariate Data Description}\label{app:user_cov}
Table \ref{tab:usercovariates} presents all the user covariate data used in our empirical analysis in Section \ref{sec:empirical_setting}.
\begin{table}[!t]
    \centering\scriptsize
    \caption{User Covariates Used in the Empirical Analysis }
    \begin{tabular}{lll}
    \hline 
    \hline \\[-1.8ex] 
    & Variable & Description\\
    \hline\\[-1.8ex] 
   \multirow{15}{0.8cm}{Disc-rete Var.}
   & Gender & Gender of user: male, female, or unknown\\
    &User Activeness Degree & Activeness of user: high-, mid-, low-active, or new user\\
    &User Life Time Stage & Life cycle phase of user on the platform: new, mature, or recession\\
    &Operating System & OS of user's device: Android, iPhone, IPAD, or other\\
    &Installation of video-sharing platform A & Installed or not\\
    &Installation of longer video-sharing platform B & Installed or not\\
    &Installation of live-streaming platform C & Installed or not\\
    &Installation of game live-streaming platform D & Installed or not\\
    &Installation of game live-streaming platform E & Installed or not\\
    &Frequent Residence Area & Region in which the user is frequently on the platform:\\
    &&South, North, or unknown\\
    &Frequent Residence City Level & Level of the city in which the user is frequently on the platform: \\
    & & large city, big city, medium city, small city, or unknown \\
    &Number of Mutual Followers & Interval of the user's number of mutual followers (friends): \\
    &&$<$5, 5 - 30, 30 - 60, 60-120, 120-250 $>$250\\
    \hline\\[-1.8ex] 
    \multirow{52}{0.8cm}{Conti-nuous Var.}
    &Age & Age of the user\\
    & Number of Followed Users & Number of users followed by the user before the treatment\\
    & Number of Fans & Number of users following the user before the treatment\\
    & Model Price& The price of the device model used by the user\\
    \cline{2-3}\\[-1.8ex] 
    & \multicolumn{2}{c}{In past 30 days on simplified app version}\\
    \cline{2-3}\\[-1.8ex] 
    &Average App Usage Duration  & User's average usage duration on platform per day\\
    & Average Video Watching Time& User's average time on watching videos on platform per day\\
    & Average Profile Page Staying Time& User's average time staying on his own profile page per day\\
    &Number of videos being completely watched & Total number of fully viewed videos by the user\\
    &Number of videos being validly watched & Total number of videos thoroughly and validly viewed by the user\\
    & Number of videos uploading & Total number of videos posted by the user\\
    &Number of app launch times  uploading & Total number of times of user launching the app\\
    \cline{2-3}\\[-1.8ex] 
    & \multicolumn{2}{c}{In past 10 days on simplified app version}\\
    \cline{2-3}\\[-1.8ex] 
    &Average App Usage Duration  & User's average usage duration on platform per day\\
    & Average Video Watching Time& User's average time on watching videos on platform per day\\
    & Average Profile Page Staying Time& User's average time staying on his own profile page per day\\
    &Number of videos being completely watched & Total number of fully viewed videos by the user\\
    &Number of videos being validly watched & Total number of videos thoroughly and validly viewed by the user\\
    & Number of videos uploading & Total number of videos posted by the user\\
    &Number of app launch times  uploading & Total number of times of user launching the app\\
        \cline{2-3}\\[-1.8ex] 
    & \multicolumn{2}{c}{In past 3 days on simplified app version}\\
    \cline{2-3}\\[-1.8ex] 
    &Average App Usage Duration  & User's average usage duration on platform per day\\
    & Average Video Watching Time& User's average time on watching videos on platform per day\\
    & Average Profile Page Staying Time& User's average time staying on his own profile page per day\\
    &Number of videos being completely watched & Total number of fully viewed videos by the user\\
    &Number of videos being validly watched & Total number of videos thoroughly and validly viewed by the user\\
    & Number of videos uploading & Total number of videos posted by the user\\
    &Number of app launch times  uploading & Total number of times of user launching the app\\
    \cline{2-3}\\[-1.8ex] 
    & \multicolumn{2}{c}{In past 30 days on original app version}\\
    \cline{2-3}\\[-1.8ex] 
    &Average App Usage Duration  & User's average usage duration on platform per day\\
    & Average Video Watching Time& User's average time on watching videos on platform per day\\
    & Average Profile Page Staying Time& User's average time staying on his own profile page per day\\
    &Number of videos being completely watched & Total number of fully viewed videos by the user\\
    &Number of videos being validly watched & Total number of videos thoroughly and validly viewed by the user\\
    & Number of videos uploading & Total number of videos posted by the user\\
    &Number of app launch times  uploading & Total number of times of user launching the app\\
        \cline{2-3}\\[-1.8ex] 
    & \multicolumn{2}{c}{In past 10 days on original app version}\\
    \cline{2-3}\\[-1.8ex] 
    &Average App Usage Duration  & User's average usage duration on platform per day\\
    & Average Video Watching Time& User's average time on watching videos on platform per day\\
    & Average Profile Page Staying Time& User's average time staying on his own profile page per day\\
    &Number of videos being completely watched & Total number of fully viewed videos by the user\\
    &Number of videos being validly watched & Total number of videos thoroughly and validly viewed by the user\\
    & Number of videos uploading & Total number of videos posted by the user\\
    &Number of app launch times  uploading & Total number of times of user launching the app\\
        \cline{2-3}\\[-1.8ex] 
    & \multicolumn{2}{c}{In past 3 days on original app version}\\
    \cline{2-3}\\[-1.8ex] 
    &Average App Usage Duration  & User's average usage duration on platform per day\\
    & Average Video Watching Time& User's average time on watching videos on platform per day\\
    & Average Profile Page Staying Time& User's average time staying on his own profile page per day\\
    &Number of videos being completely watched & Total number of fully viewed videos by the user\\
    &Number of videos being validly watched & Total number of videos thoroughly and validly viewed by the user\\
    & Number of videos uploading & Total number of videos posted by the user\\
    &Number of app launch times  uploading & Total number of times of user launching the app\\
    \hline 
    \hline \\[-1.8ex] 
    \end{tabular}
    \label{tab:usercovariates}
\end{table}
\subsection{Implementation {\sf DLPT} on Real-World Experiment Data}\label{app:empirical_implementation}
In this section, we introduce the details of the implementation of {\sf DLPT} on real-world experiment data. We follow the three-stage approach outlined in Section \ref{sec:theo_framework}. First, we implement the assumed DGP using a structured deep learning model. This allows us to obtain estimators of nuisance parameters. Second, we estimate the average treatment effect and policy value through the doubly robust estimators. Finally, we proceed to optimize the policy value by finding the policy with minimal regrets, aiming to identify the most effective policy for enhancing user video posting behavior while balancing costs and benefits. 

In Stage 1, we assume that the outcome $Y$, user features $\bm{X}$, and continuous treatment level $T$ (rescale by diving $1,000$ for better training) follow the DGP specified in Equation (\ref{eq:G_func}) where we set polynomial degree $K=3$.
\begin{equation}\label{eq:G_func}
    \mathbb{E}[Y | \bm X = \bm x, T= t]=G(\bm{\theta}_K^*(\bm{x}),t) = \frac{1}{1+\exp(-(\theta^*_0(\bm x)+\theta^*_1(\bm x)t +\theta^*_2(\bm x)t^2+ \theta^*_3(\bm x)t^3))}.
\end{equation}
This DGP structure captures a non-linear response to treatment levels, effectively modeling diminishing returns as treatment intensity increases. Moreover, it provides a nuanced understanding of how variations in the treatment variable influence outcomes across diverse individual features, ensuring both flexibility and robustness in the analysis.

The policy value for a given policy $\pi$ is specified as follows:
\begin{equation*}\label{eq:policy_value_empirical}
    V(\pi)=\bbE[wY-c\pi(\bm X)],
\end{equation*}
where $w=0.5$ and $c=0.1$ are parameters calculated by Platform O.

To implement this DGP, we use a structured DNN illustrated in Figure \ref{fig:dnn_structure}. The structured DNN takes covariates (with a dimension equal to 77) and treatment (with a dimension equal to 1) as input. The covariates go through a three-layer DNN with 128 nodes, 32 nodes, and 8 nodes, and subsequently the approximated parameters layer $\theta^*_k(\cdot)$, for $k\in\{0,1,2,3\}$. We use the $ReLU$ function for each layer such that $ReLU(x)=\max\{0,x\}$ as the activation function. Treatment input $t$ then concatenate with nuisance parameters $\theta_k(\cdot)$ for $k\in\{1,2,3\}$ with corresponding degree. Taking the linear additivity of $\theta_0(\bm x)$ and $\theta_k(\bm x)t^k$ as the input of the sigmoid function, we then obtain the outcome $y$. The structured DNN is implemented via Tensorflow.

The training dataset $\mathcal{S}_{\text{train}}$ is used to fit in the structured DNN to obtain estimators of nuisance parameters $\hat{\theta}_k$ for $k\in\{0,1,2,3\}$, which is trained under a binary cross entropy loss function through the ADAM optimizer \citep{kingma2014adam}. We set the learning rate $=0.0001$, batch size $=128$, and leave $10\%$ of the training data  $\mathcal{S}_{train}$ as a validation dataset for better training results. After 100 epochs of training, we obtain the estimator $\hat{\theta}_k(\cdot)$ for $k\in\{0,1,2,3\}$.

We continue to the second stage, where we construct estimators for policy value for each level of points as defined in Equation \eqref{eq:dr_estimator} using the inference dataset  $\mathcal{S}_{\text{inference}}$. 

In the last stage, we derive the personalized optimal policies by solving the ERM problem $ \hat{\pi}_{DLPT} = \inf_{\pi\in{\Pi}}~\{-\hat{V}_K^{DLPT}(\pi)\}$ using the inference dataset $\mathcal{S}_{\text{inference}}$. 
For discrete policy classes, including the observed discrete policy class and the discretized continuous policy class, we compute the Neyman orthogonal policy value $\hat{V}_K^{DLPT}(\pi)$ for each candidate action (arm) and select the arm with the highest estimated value. For continuous policy classes, including linear policy class and DNN policy class, we adopt $-\hat{V}_K^{DLPT}(\pi)$ as the loss function and utilize TensorFlow's automatic differentiation to compute gradients, optimizing the policy $\pi$ for a given policy class using the Adam optimizer with a learning rate of 0.001. Early stopping is applied with a patience parameter of 5 to prevent overfitting.
\subsection{Implementation Details of Benchmark Methods on Real-World Experiment Data}\label{app:imple_bench_real}
In this section, we document the implementation details of each benchmark method on real-world experiment data.
 All benchmarks use the same training dataset $\mathcal{S}_{\text{train}}$ as used for  {\sf DLPT} and share the same inference dataset $\mathcal{S}_{\text{inference}}$ with  {\sf DLPT}. For the initial four benchmark methods, ${{\sf LR}, {\sf LogR}, {\sf LCM}, {\sf PDL}}$, the objective is to perform both ATE estimation and policy learning by hypothesizing various DGPs for the outcome $Y$ based on user features $\bm{X}$ and the continuous treatment $T$.

\textbf{{\sf LR}} utilizes user covariates $\bm X$ and the points treatment $t$, operating under the assumption that the likelihood of user $i$ uploading a public video adheres to the linear regression model specification: $y_i = \bm{\alpha'} \bm{x_i} + \beta t_i +\mu_i$. Here, $\hat{\beta}$, the estimator of the coefficient of $t_i$, quantifies the boosting effect of the points on users' video-posting behavior.
In the first estimation stage, we fit this linear regression model using the data points from $\mathcal{S}_{\text{train}}$. For the second stage of inference, we predict the video-uploading outcomes for users in $\mathcal{S}_{\text{inference}}$ across four point levels. We then calculate the ATE estimators for the four different point policies by conducting pair-wise t-tests on these predicted outcomes. In the last stage of policy learning, we calculate the estimated policy value of four point policies within one subgroup $j$, and then identify the optimal policy by selecting the one with the highest estimated policy value.

\textbf{{\sf LogR}} transforms the outcome of the linear regression model from a continuous variable to a categorical variable. Leveraging user $i$ feature $\bm{x_i}$ and continuous treatment $t_i$, the logistic regression model assumes that: $\mathbb{E}[y_i]=\frac{1}{1+\exp({-( \bm{\alpha'} \bm{x_i} + \beta t_i)})}$. Similar to the three-stage process in {\sf LR}, we use $\mathcal{S}_{\text{train}}$ for training and $\mathcal{S}_{\text{inference}}$ for ATE inference and policy learning.

\textbf{{\sf LCM}} stems from the traditional economics literature. Users face two choices: upload videos or not. We assume that user $i$ has an individual utility over the point policy such that $u_i = \bm{\alpha'} \bm{x_i} + \bm\beta \bm{x_i} t_i +\epsilon_i$, where $\epsilon_i$ follows a Gumbel distribution. Without loss of generality, we normalize the utility of not uploading videos to zero. The probability of user $i$ choose to upload videos follows the logit choice model $y_i=\frac{\exp(u_i)}{1+\exp(u_i)}$. To fit the choice model, we employ the maximum likelihood estimation (MLE) method under the ``l-bfgs-b'' algorithm \citep{zhu1997algorithm} using $\mathcal{S}_{\text{train}}$. We then follow the similar logic as {\sf LR} and {\sf LogR} to obtain the predicted outcome for users in $\mathcal{S}_{\text{inference}}$ and conduct pair-wise t-tests for ATE estimators as well as optimal policy identification.

\textbf{{\sf PDL}} employs fully flexible DGP assumption such that outcome $y_i=f(\bm x_i,t_i)$. To maintain comparability, we use a DNN with similar structure as illustrated in Figure \ref{fig:dnn_structure} to approximate function $f(\cdot)$. The difference of DNN between {\sf PDL} and {\sf DLPT} is that unlike {\sf DLPT}, both covariates $\bm x_i$ and treatment $t_i$ in {\sf PDL} go through 3 hidden layers and then go straight to the last layer as the input of a sigmoid function. We use the same hyperparameter set as {\sf DLPT} implementation in Section \ref{sec:empirical_application} such that \{learning rate $=0.0001$, batch size $=128$, validation proportion $=0.1$\}. The {\sf PDL} approach follows the same pair-wise t-test as benchmarks {\sf LR}, {\sf LogR}, and {\sf LCM} based on the predicted outcome given users in $\mathcal{S}_{\text{inference}}$. We remark that {\sf PDL} offers the greatest flexibility, imposing no restrictions on the functional form for DGP assumption while leveraging the maximum predictive power of DNN. However, {\sf PDL} cannot utilize the influence function to debias the estimates.


\textbf{{\sf UNI}} assigns a non-personalized policy to all users. Specifically, we choose the point policy with the highest $\hat{V}_{\sf UNI}(\pi) = \frac{w}{|\mathcal{S}_{\text{train}},t_i=\pi|}\sum_{i\in \mathcal{S}_{\text{train}},t_i=\pi}y_i - c\pi$, where $|\mathcal{S}_{\text{train}},t_i=\pi|$ is the number of users in $\mathcal{S}_{\text{train}}$ with observed treatment equals to $\pi$.


\section{Robustness Checks with Semi-Synthetic Studies}\label{app:synthetic}

 \subsection{Robustness to Treatment Level Coverage}\label{sec:robust to treatment}
In this section, we assess the robustness of the {\sf DLPT} framework concerning the coverage range of the tested treatment levels. According to Proposition \ref{prop: nuisance convergence_rate}, a minimum of four distinct levels of treatment, $t$, must be randomly assigned to collect data $\bm Z$ for implementing the {\sf DLPT} framework. Notably, while our framework stipulates no restrictions on the coverage range of $t$, practical implementations often operate within a limited range. For instance, even though the continuous policy $\pi(\bm x)$ might span from $[0,1]$, in reality, the treatment levels tested could be as narrow as $\{0,0.068,0.128,0.168,0.25\}$. This scenario underscores the need to evaluate the framework’s performance under constrained experimental designs and to understand how such limitations might affect the generalizability and effectiveness of the policy decisions derived from these data. In Figure \ref{fig:robustness}, we report the policy regret results under four different treatment level coverage: (1) $[0,0.25]$ with treated level $t\in\{0,0.068,0.128,0.168,0.25\}$. (2)$[0,0.50]$ with treated level $t\in\{0,0.128,0.268,0.358,0.5\}$. (3) $[0,0.75]$ with treated level $t\in\{0,0.178,0.358,0.538,0.75\}$. (2)$[0,1.00]$ with treated level $t\in\{0,0.268,0.538,0.718,1\}$.

 Our analysis reveals that broader treatment coverage contributes to improved policy learning outcomes; however, the incremental gains diminish as the range extends from $[0,0.75]$ to $[0,1.00]$. Managerially, this suggests that slightly reducing treatment coverage in real-world experimental settings could effectively lower experiment costs without significantly compromising the results, as higher levels of treatment typically entail increased expenses. The moderate robustness of the {\sf DLPT} framework regarding treatment coverage offers valuable insights for optimizing future experimental designs.

 \subsection{Regret over Assumed Polynomial Degree}
In this section, we evaluate the performance of the {\sf DLPT} framework under varying assumed polynomial degrees \( K \in \{1, 2, 3, 4\} \). As illustrated in the right panel of Figure~\ref{fig:robustness}, policy regret decreases as \( K \) increases, reflecting the improved approximation power of higher-order models. However, the incremental improvement tapers off between \( K = 3 \) and \( K = 4 \). This diminishing gain is likely due to the increased estimation complexity: a higher \( K \) introduces an additional nuisance parameter that must be learned. In practice, this imposes a greater burden on the structured DNN, making convergence to the true optimal nuisance function more challenging given the fixed training data size.

\end{APPENDICES}
\end{document}